\documentclass[12pt]{article}
\pdfoutput=1
\usepackage{epsf,amsmath,amssymb,graphicx,dcolumn}
\usepackage{scalefnt,cite,hyperref}
\usepackage{cleveref,etoolbox,color,soul}
\usepackage{listings}
\usepackage[utf8]{inputenc}

\Crefname{figure}{Fig.}{Figs.}
\usepackage{placeins}
\usepackage{soul}

\newcommand{\citere}[1]{Ref.\,\cite{#1}}
\newcommand{\citeres}[1]{Refs.\,\cite{#1}}
\newcommand{\code}{\tt}

\newcommand{\fhversion}{{\tt 2.14.3}}
\newcommand{\sushi}[1]{{\code SusHi#1}}
\newcommand{\FH}{{\code FeynHiggs}}
\newcommand{\CPsuperH}{{\code CPsuperH}}

\newcommand{\abbrev}{\scalefont{1}}
\newcommand{\eqn}[1]{Eq.\,(\ref{#1})}
\newcommand{\fig}[1]{Fig.\,\ref{#1}}
\newcommand{\figs}[1]{Figs.\,\ref{#1}}

\newcommand{\lhc}{{\abbrev LHC}}

\newcommand{\sm}{{\abbrev SM}}
\newcommand{\thdm}{{\abbrev 2HDM}}
\newcommand{\mssm}{{\abbrev MSSM}}

\newcommand{\susy}{{\abbrev SUSY}}
\newcommand{\bsm}{{\abbrev BSM}}

\newcommand{\EW}{{\abbrev EW}}
\newcommand{\atlas}{{\abbrev ATLAS}}
\newcommand{\cms}{{\abbrev CMS}}
\newcommand{\lep}{{\abbrev LEP}}
\newcommand{\lhchxswg}{{\abbrev LHC-HXSWG}}

\newcommand{\cp}{{\abbrev $\mathcal{CP}$}}
\newcommand{\nlo}{{\abbrev NLO}}
\newcommand{\nnlo}{{\abbrev NNLO}}
\newcommand{\nklo}[1]{{\abbrev N$^{#1}$LO}}
\newcommand{\drbar}{{\abbrev $\overline{\text{DR}}$}}

\newcommand{\muF}{\mu_\text{F}}
\newcommand{\muR}{\mu_\text{R}}

\newcommand{\mW}{\ensuremath{M_W}}
\newcommand{\mZ}{\ensuremath{M_Z}}
\newcommand{\mh}{\ensuremath{M_h}}
\newcommand{\mH}{\ensuremath{M_H}}
\newcommand{\mA}{\ensuremath{M_A}}
\newcommand{\mhp}{\ensuremath{M_{H^\pm}}}
\newcommand{\tb}{\ensuremath{\tan\beta}}
\newcommand\ino[1]{\tilde\chi_{#1}}

\newcommand\neu[1]{\ino{#1}^0}

\definecolor{darkgreen}{rgb}{0,0.5,0.1}

\newcommand{\mhsc}{$\mh^\text{125}$}
\newcommand{\lstau}{$\mh^{125}(\tilde{\tau})$}
\newcommand{\lchi}{$\mh^{125}(\tilde{\chi})$}
\newcommand{\halignment}{$\mh^\text{125}(\mathrm{alignment})$}
\newcommand{\mHsc}{$\mH^\text{125}$}
\newcommand{\cpvint}{$M_{h_1}^{\text{125}}(\text{CPV})$}

\newcounter{notecount}
\begin{document}
\thispagestyle{empty}
\vspace*{-1.8cm}
\begin{flushright}
{\tt MPP-2018-211},
{\tt DESY 18-140},
{\tt KA-TP-25-2018},\\
{\tt IFT-UAM/CSIC-18-017},
{\tt EFI-18-12},
{\tt PSI-PR-19-13}
\end{flushright}

\long\def\symbolfootnote[#1]#2{\begingroup%
\def\thefootnote{\fnsymbol{footnote}}\footnote[#1]{#2}\endgroup}

\vspace{1cm}

\begin{center}
\Large\bf\boldmath
MSSM Higgs Boson Searches at the LHC:\\[0.2cm]
Benchmark Scenarios for Run 2 and {B}eyond
\unboldmath
\end{center}
\vspace{0.05cm}
\begin{center}
{Emanuele Bagnaschi$^a$,}  
Henning Bahl$^b$,
Elina Fuchs$^c$,
Thomas Hahn$^b$,\\[0.2cm]
Sven Heinemeyer$^{d,e,f}$,
Stefan Liebler$^g$,
Shruti Patel$^{g,h}$,
Pietro Slavich$^{i}$,\\[0.2cm]
Tim Stefaniak$^{j}$,
Carlos~E.M.~Wagner$^{k,l,m}$ and
Georg Weiglein$^j$\\[0.4cm]
{\small
{{\sl ${}^a$ Paul Scherrer Institut, CH-5232 Villigen PSI, Switzerland}}\\[0.2em]
  {\sl${}^b$Max-Planck Institut f\"ur Physik, D-80805 M\"unchen, Germany}\\[0.2em]  
{\sl${}^c$Department of Particle Physics and Astrophysics,\\
Weizmann Institute of Science, Rehovot 76100, Israel}\\[0.2em]
{\sl${}^d$Instituto de F\'isica de Cantabria (CSIC-UC), E-39005 Santander, Spain}\\[0.2em]
{\sl${}^e$Instituto de F\'isica Te\'orica, (UAM/CSIC), Universidad Aut\'onoma de Madrid,\\
Cantoblanco, E-28049 Madrid, Spain}\\[0.2em]
{\sl${}^f$Campus of International Excellence UAM+CSIC, Cantoblanco, E-28049, Madrid, Spain}\\[0.2em]
{\sl${}^g$Institute for Theoretical Physics (ITP), Karlsruhe Institute of Technology,\\ D-76131 Karlsruhe, Germany}\\[0.2em]
{\sl${}^h$Institute for Nuclear Physics (IKP), Karlsruhe Institute of Technology,\\ D-76344 Karlsruhe, Germany}\\[0.2em]
{\sl${}^i$Sorbonne Universit\'e, CNRS, Laboratoire de Physique Th\'eorique et Hautes \'Energies, LPTHE, F-75005 Paris, France}\\[0.2em]
{\sl${}^j$DESY, Notkestra{\ss}e 85, D-22607 Hamburg, Germany}\\[0.2em]
{\sl${}^k$High Energy Physics Division, Argonne National Laboratory, Argonne, IL 60439, USA}\\[0.2em]
{\sl${}^l$Enrico Fermi Institute, University of Chicago, Chicago, IL 60637, USA}\\[0.2em]
{\sl${}^m$Kavli Institute for Cosmological Physics, University of Chicago, Chicago, IL 60637, USA}
}
\end{center}
\vspace*{1mm}
\begin{abstract}
\noindent
We propose six new benchmark scenarios for Higgs boson searches in the
Minimal Supersymmetric Standard Model. Our calculations follow the
recommendations of the \lhc\ Higgs Cross Section Working Group, and
benefit from recent developments in the {predictions for} the
Higgs-boson masses and mixing.  All of the proposed scenarios are
compatible with the most recent results from Run 2 of the LHC. In
particular, they feature a scalar with mass and couplings compatible
with those of the observed Higgs boson, and a significant portion of
their parameter space is {allowed by the limits from the} searches
for SUSY particles and additional Higgs bosons.
We define a scenario where all SUSY particles are relatively heavy,
and two scenarios with light colorless SUSY particles (charginos,
neutralinos and, in one case, staus).  In addition, we present two
scenarios featuring alignment without decoupling, realized with either
the lighter or the heavier scalar being SM-like, and a scenario with
\cp\ violation.

\end{abstract}

\setcounter{footnote}{0}

\newpage

\pagenumbering{arabic}

\section{Introduction}
\label{sec:introduction}

In the six years since the discovery of a Higgs boson at the
LHC~\cite{Aad:2012tfa,Chatrchyan:2012xdj}, its mass has been measured
with a few-per-mil accuracy, $M_H^{\rm obs} = 125.09 \pm
0.24$~GeV~\cite{Aad:2015zhl}.  The measured properties are, within
current experimental and theoretical uncertainties, in agreement with
the predictions of the Standard Model (SM)~\cite{Khachatryan:2016vau}.
Together with the current lack of discoveries of beyond-the-SM (BSM)
particles at the LHC with center-of-mass energies of up to 13~TeV, the
requirement that the particle spectrum include an essentially SM-like
Higgs boson {imposes} important constraints on the parameter space
of possible extensions of the SM.

The Minimal Supersymmetric Standard Model
(MSSM)~\cite{Nilles:1983ge,Haber:1984rc,Gunion:1984yn} is one of the
best motivated among those extensions -- and probably the most
studied. Its Higgs sector consists of two SU(2) doublets, $H_1$ and
$H_2$, whose relative contribution to electroweak symmetry breaking
(EWSB) is determined by the ratio of vacuum expectation values (vevs)
of their neutral components, $\tan\beta \equiv v_2/v_1$.  At the tree
level, \cp\ is conserved in the Higgs sector of the MSSM, and the
spectrum of physical Higgs bosons consists of two neutral \cp-even
scalars, of which we denote the lighter as $h$ and the heavier as $H$,
one \cp-odd scalar, $A$, and a charged-scalar pair, $H^\pm$.
Supersymmetry (SUSY) imposes relations between the quartic Higgs
couplings and the gauge couplings, ensuring that the tree-level masses
of all Higgs bosons can be expressed in terms of the gauge-boson
masses, $\mZ$ and $\mW$, plus two additional parameters which can be
chosen as the \cp-odd scalar mass, $\mA$ (or alternatively the charged
Higgs boson mass, \mhp), and $\tan\beta\,$. In particular, the
tree-level mass of the lighter \cp-even scalar $h$ is bounded from
above by $\mZ\,|\cos2\beta|$. However, radiative corrections --
especially those involving top and bottom quarks and their scalar
partners, the stops and the sbottoms -- can significantly alter the
tree-level predictions for the Higgs masses, allowing for $\mh \approx
125$~GeV but bringing along a dependence on many free parameters of
the MSSM, see \citere{Draper:2016pys} for a recent review.  Moreover,
for specific choices of those parameters, radiative corrections to the
mixing between the scalars can also allow for scenarios in which the
heavier mass eigenstate, $H$, is the one with $\mH \approx 125$~GeV
and roughly SM-like couplings, see e.g.~\citeres{Heinemeyer:2011aa,
  Bechtle:2012jw, Bechtle:2016kui, Haber:2017erd}.
{In} the presence of complex parameters in the MSSM Lagrangian,
radiative corrections can break \cp\ in the Higgs sector and induce a
mixing among the two \cp-even scalars, $h$ and $H$, and the \cp-odd
scalar, $A$, such that beyond tree-level they combine into three
neutral mass eigenstates which we denote as $h_a$ (with $a=1,2,3$).

The large number of free parameters complicates the task of
interpreting within the MSSM both the properties of the observed Higgs
boson and the results of the ongoing searches for additional,
non-standard Higgs bosons. Complete scans of the MSSM parameter space
would be highly impractical for experimental analyses and
phenomenological studies. Therefore, a number of {\it benchmark
  scenarios} has been proposed over the years, for both the
\cp-conserving~\cite{Carena:1999xa, Carena:2002qg, Carena:2013ytb} and
\cp-violating~\cite{Carena:2000ks,Carena:2015uoe} cases. In these
scenarios, two parameters in the Higgs sector are varied -- typically,
one of them is $\tan\beta$ and the other is either $\mA$, for the
\cp-conserving case, or $\mhp$, for the \cp-violating case -- while
the remaining parameters (such as the soft{-}SUSY-breaking masses and
mixing terms for the sfermions, as well as the masses of gauginos and
higgsinos) are fixed to values chosen to exhibit certain aspects of
MSSM Higgs phenomenology. In particular, \citere{Carena:2013ytb}
proposed seven \cp-conserving benchmark scenarios that, over a wide
range of values of the two free parameters, featured a Higgs boson
whose properties were compatible with those measured during Run 1 of
the LHC. Three of those scenarios involved relatively heavy colored
superpartners, with masses of the order of one TeV, and lighter
charginos and neutralinos with masses around $200$~GeV or less; one
scenario involved a light stop of mass about $325$~GeV, with a
``compressed'' stop{--}neutralino spectrum to avoid the exclusion
bounds from direct stop searches; one scenario involved a light stau,
allowing for a sizable contribution to the diphoton width of the
SM-like Higgs boson; in one scenario the radiative corrections to the
Higgs mixing led to reduced couplings to down-type fermions; finally,
in the last scenario the heavier \cp-even scalar $H$ could be
identified with the observed Higgs boson.

The LHC Higgs Cross Section Working Group (\lhchxswg) produced a set
of {\tt ROOT} files~\cite{Heinemeyer:2013tqa,deFlorian:2016spz}
providing, for each of the benchmark scenarios of
\citere{Carena:2013ytb}, what were then state-of-the-art predictions
for the masses, production cross sections and decay widths of the
three neutral Higgs bosons, over a grid of values of $\mA$ and
$\tan\beta$ (except for the last scenario, in which the free
parameters are $\tan\beta$ and the Higgs/higgsino superpotential mass
$\mu\,$). Those predictions were subsequently used by both the ATLAS
and CMS
collaborations to interpret the results of their searches for
additional scalars in the context of the MSSM.

Important developments in the years since the publication of
\citere{Carena:2013ytb} motivate a reassessment of the benchmark
scenarios presented there. On the one hand, the full analysis of Run-1
LHC data at center-of-mass energies of $7$ and $8$~TeV, as well as
{the available} analyses of Run-2 data at $13$~TeV, have tightened the
experimental constraints on masses and couplings of both the observed
Higgs boson and any still-unobserved \bsm\ particles. On the other
hand, the theoretical predictions for the MSSM Higgs-boson masses have
evolved: the renewed interest in SUSY scenarios with heavy
superpartners (i.e., with masses larger than a few TeV) has stimulated
new calculations~\cite{Hahn:2013ria,Draper:2013oza,Bagnaschi:2014rsa,
  Vega:2015fna,Bahl:2016brp,Athron:2016fuq,Staub:2017jnp,Bagnaschi:2017xid,
  Bahl:2017aev, Allanach:2018fif, Harlander:2018yhj, KUTS} aiming at
the resummation of potentially large corrections enhanced by
logarithms of the ratio between the SUSY scale and the EWSB scale. The
versions of the code \FH~\cite{Heinemeyer:1998yj} used both to devise
the scenarios of \citere{Carena:2013ytb} and in the production of the
corresponding \lhchxswg\ files relied on a fixed-order calculation of
the MSSM Higgs masses, employing full one-loop corrections from
\citere{Frank:2006yh} and the dominant two-loop corrections from
\citeres{Heinemeyer:1998np, Degrassi:2001yf, Brignole:2001jy,
  Brignole:2002bz, Dedes:2003km}. In contrast, newer versions of
\FH\ include also a full next-to-leading-logarithmic (NLL) and partial
next-to-NLL (NNLL) resummation of the large logarithmic
corrections~\cite{Hahn:2013ria, Bahl:2016brp, Bahl:2017aev}, based on
an effective-field-theory (EFT) approach. Implementing this
resummation required modifications in the code that, even for the stop
masses around one TeV featured in the scenarios of
\citere{Carena:2013ytb}, can lower the prediction for $\mh$ by
       {$1\!-\!2$~GeV}.
For an example of the combined effect of these developments, we can
consider the case of the ``light-stop'' scenario of
\citere{Carena:2013ytb}: pair production of the lighter stop followed
by the decay $\tilde t_1 \to c\,\neu1$ is now excluded by monojet
searches by ATLAS~\cite{Aaboud:2017phn} and
CMS~\cite{Sirunyan:2017kiw} for stop masses of up to $430$~GeV and
$510$~GeV, respectively, unless the lightest neutralino is so close in
mass to the stop that the latter becomes long-lived.  In addition, the
lowering of the prediction for $\mh$ in recent versions of \FH\ makes
the ``light-stop'' scenario of \citere{Carena:2013ytb} incompatible
with an observed Higgs mass of about $125$~GeV, even within a
theoretical uncertainty usually estimated as $\pm
3$~GeV~\cite{Degrassi:2002fi,Allanach:2004rh}.

For the MSSM with complex parameters, and thus a mixing among all
three neutral Higgs states, benchmark scenarios were originally
proposed in \citere{Carena:2000ks}, to be used in the interpretation
of the searches for MSSM Higgs bosons at
LEP~\cite{Schael:2006cr}. They were later updated in
\citere{Carena:2015uoe} to include a \sm-like Higgs boson with mass
around $125$~GeV.  In those papers, the masses and couplings of the
Higgs bosons were computed with the code \CPsuperH~\cite{Lee:2003nta,
  Lee:2007gn, Lee:2012wa}{, relying on the calculation of
  \citeres{Pilaftsis:1999qt, Choi:2000wz, Carena:2000yi}}. However, no
scenario with \cp\ violation in the Higgs sector has been considered
so far in the context of the \lhchxswg.
A reassessment of the benchmark scenarios for the MSSM with complex
parameters seems now well motivated, in view of tightening constraints
from electric dipole moments (EDMs)~\cite{Baker:2006ts,
  Baron:2013eja}, as well as of recent progress in the calculation of
the Higgs masses (allowing for complex phases in the dominant two-loop
corrections~\cite{Heinemeyer:2007aq, Hollik:2014wea, Hollik:2014bua,
  Hollik:2015ema, Passehr:2017ufr, Borowka:2018anu} and,
approximately, in the large-log resummation~\cite{Bahl:2017aev}), in
the prediction of the cross sections for Higgs
production~\cite{Liebler:2016ceh} and in the incorporation of
interference effects among admixed Higgs bosons~\cite{Fuchs:2014ola,
  Fuchs:2016swt, Fuchs:2017wkq}. 
 
Another type of MSSM benchmark scenarios that was studied by the
\lhchxswg~\cite{Bagnaschi:2015hka} is characterized by very low values
of $\tan\beta$, and requires very heavy SUSY particles.  Indeed, the
tree-level \mssm\ prediction for $\mh$ goes to zero as $\tan\beta$
approaches unity, thus, at very low $\tan\beta$, stop masses much
larger than a few TeV are needed to obtain $\mh \approx 125$~GeV
through radiative corrections.
The precise calculation of the Higgs masses and couplings in these
scenarios, including a proper resummation of large logarithmic
corrections, requires an EFT approach in which the theory valid below
the SUSY scale is a two-Higgs-doublet model (\thdm). Work in this
direction has been presented in \citeres{Lee:2015uza,
  Bagnaschi:2015pwa}, and the setup with heavy SUSY and a light
\thdm\ is also being implemented in \FH~\cite{Bahl:2018jom}. In this
paper, however, we will focus on benchmark scenarios with SUSY
particles around the TeV scale, in which the region with very low
$\tan\beta$ is simply ruled out by an excessively low prediction for
the mass of the SM-like Higgs boson.

We finally mention that, in some instances,
the ATLAS and CMS collaborations relied on a simplifying approach, the
so-called ``hMSSM''~\cite{Djouadi:2013vqa, Maiani:2013hud,
  Djouadi:2013uqa, Djouadi:2015jea}, to interpret their Higgs searches
in the context of the MSSM. This approximation assumes that the Higgs
sector is \cp{} conserving, that all superparticles are too heavy to
affect Higgs production and decays, that any non-decoupling SUSY
corrections to the Higgs couplings are negligible, and that the
radiative corrections to the elements other than (2,2) in the mass
matrix of the neutral \cp-even components of $H_1$ and $H_2$ are also
negligible, i.e.~$\Delta {\cal M}_{1j}^2 \approx 0$ for $j=1,2$. In
that case, the remaining radiative correction $\Delta {\cal M}_{22}^2$
can be expressed in terms of the parameters that determine the
tree-level mass matrix (i.e.~$\tan\beta$, $\mZ$ and $\mA$) plus the
smaller eigenvalue $\mh$, which is treated as an input and identified
with the mass of the observed Higgs boson. Consequently, the larger
eigenvalue $\mH$, the mixing angle $\alpha$ and the Higgs
self-couplings can in turn be expressed in terms of just those four
input parameters, of which only $\tan\beta$ and $\mA$ are unknown.
While the hMSSM approach brings some clear benefits -- namely, the
limited number of input parameters and the fact that the condition
$\mh = 125.09$~GeV can be enforced all over the $(\mA,\tan\beta)$
plane -- its predictions for the Higgs properties can be mapped only
to regions of the \mssm\ parameter space in which the approximations
of neglecting the $\Delta {\cal M}_{1j}^2$ corrections and the SUSY
corrections to the Higgs couplings are justified. That, however, is
not necessarily the case for small values of $\mA$ and for rather
large values of $\mu$ and $\tan\beta$. Moreover, near the
{lower}-left corner of the $(\mA,\tan\beta)$ plane -- i.e., for
$\tan\beta \lesssim 1.5$ and $\mA \lesssim 200$~GeV -- a
lighter-scalar mass of about $125$~GeV may require that the stops are
as heavy as the GUT scale, putting into question the validity of the
MSSM as the underlying high-energy theory~\cite{Lee:2015uza}. In
summary, an unambiguous interpretation of the results of the Higgs
searches within the \mssm\ cannot really elude the dependence of the
Higgs masses and couplings on the underlying SUSY parameters, and
requires the definition of appropriate benchmark scenarios.

\bigskip 

In this paper we propose six new benchmark scenarios for MSSM Higgs
searches that are compatible with the most recent LHC results for the
Higgs-boson properties and the bounds on masses and couplings of new
particles. The first scenario is characterized by relatively heavy
superparticles, so the Higgs phenomenology at the LHC resembles that
of a \thdm\ with MSSM-inspired Higgs couplings. The second and third
scenario are characterized by some of the superparticles (staus
{and/or} charginos and neutralinos) being lighter than the others
and affecting the Higgs decays. The fourth and fifth scenario are
characterized by the phenomenon of ``alignment without
decoupling''~\cite{Gunion:2002zf,Craig:2013hca,Carena:2013ooa,
  Carena:2014nza,Bechtle:2016kui,Haber:2017erd}, in which one of the
two neutral \cp-even Higgs scalars has SM-like couplings independently
of the mass spectrum of the remaining Higgs bosons. Here, we present
one scenario for each case, i.e.~for either $h$ or $H$ being the
SM-like scalar with mass around $125\,\mathrm{GeV}$. Finally, the
sixth scenario incorporates \cp\ violation in the Higgs sector{,
  giving rise to} a strong admixture of the two heavier neutral states
{and} leading to significant interference effects in their
production and decay.  We employ state-of-the-art calculations of the
masses, branching ratios and production cross sections of the neutral
Higgs bosons, including the effects of the resummation of large
logarithmic corrections. In particular, we use
\FH~\cite{Heinemeyer:1998yj, Heinemeyer:1998np,
  Degrassi:2002fi,Frank:2006yh, Hahn:2013ria, Bahl:2016brp,
  Bahl:2017aev} to compute the masses, mixing and branching ratios of
the neutral Higgs bosons, \sushi~\cite{Harlander:2012pb,
  Harlander:2016hcx} to compute their production cross sections, and
{\tt HiggsBounds}~\cite{Bechtle:2008jh,
  Bechtle:2011sb,Bechtle:2013wla,Bechtle:2015pma} and {\tt
  HiggsSignals}~\cite{Bechtle:2013xfa} to investigate the existing
constraints on the parameter space of the scenarios.

The rest of the paper is organized as follows: in
section~\ref{sec:theorysetup} we describe in detail the theoretical
ingredients of our calculations; in section~\ref{sec:benchmarks} we
define our six scenarios and discuss the bounds on their parameter
space; finally, section~\ref{sec:conclusions} contains our
conclusions.


\section{Theory setup}
\label{sec:theorysetup}

In this section we provide details on the precise {predictions
  for} the Higgs-boson masses, branching ratios and production cross
sections that we use to define our new benchmark scenarios. Finally,
we define the interference factors relevant to Higgs production and
decay in the MSSM with \cp\ violation.

\subsection{Higgs-boson masses and branching ratios}

For the calculation of the Higgs-boson masses we employ the latest
version of \FH, \fhversion.  While the versions used to define the
\cp-conserving scenarios of \citere{Carena:2013ytb} and to produce the
corresponding \lhchxswg\ files relied on a fixed-order (namely, full
one-loop~\cite{Frank:2006yh} and partial
two-loop~\cite{Heinemeyer:1998np, Degrassi:2001yf, Brignole:2001jy,
  Brignole:2002bz, Dedes:2003km}) calculation of the masses, the
latest version of the code includes also a full NLL and partial NNLL
resummation of the corrections involving logarithms of the ratio
between the SUSY scale and the EWSB scale~\cite{Hahn:2013ria,
  Bahl:2016brp, Bahl:2017aev}. Since our scenarios contain only
moderately heavy superparticles, with masses up to about $2.5$~TeV, we
do not expect the inclusion of three- and higher-loop logarithmic
terms to have a {very large} impact on the prediction for the
SM-like Higgs mass. However, we stress that the full NLL resummation
allows \FH\ to account -- up to terms suppressed by inverse powers of
the SUSY scale -- also for the logarithmic part of the two-loop
corrections involving the electroweak gauge couplings, which were not
included in the original fixed-order calculation. As mentioned
earlier, the combined effect of these refinements is a lower
prediction (by $1\,$--$\,2$~GeV) for the mass of the SM-like Higgs
boson in our scenarios.

For the calculation of the Higgs masses and mixing in the MSSM with
complex parameters, \FH\ includes the full one-loop results of
\citere{Frank:2006yh} and the dominant two-loop corrections involving
the top Yukawa coupling from \citeres{Heinemeyer:2007aq,
  Hollik:2014wea, Hollik:2014bua, Hollik:2015ema}.  Additional
two-loop corrections involving the bottom Yukawa coupling, as well as
the resummation of higher-order logarithmic effects, are approximated
by interpolation of the corresponding corrections computed in the MSSM
with real parameters.

The theoretical uncertainty of the prediction of \FH\ for the Higgs
masses depends on the size and the origin of the radiative
corrections, and should in principle be estimated separately {at}
each point of the MSSM parameter space. For simplicity, in the
definition of our scenarios we will stick to the global estimate of
\citeres{Degrassi:2002fi, Allanach:2004rh}, and consider all points in
the MSSM parameter space where \FH\ predicts the existence of a
SM-like Higgs boson with a mass in the window $125.09\pm3$~GeV {to
  be compatible with the Higgs-mass measurement at the LHC. This}
appears to be a conservative choice.

For future reference, we list here the values of the input flags of
\FH~\fhversion\ used in our {predictions} (see the online manual
of the code~\cite{fhsetflags} for more details):
\begin{center}
\text{mssmpart} = 4, 
\quad \text{higgsmix} = 2,\quad \text{p2approx} = 4,\quad
 \text{looplevel} = 2,\\[2mm] 
\text{loglevel} = 3,\quad \text{runningMT} = 1,
\quad \text{botResum} = 1,\quad \text{tlCplxApprox} = 0.
\end{center}
In our fifth scenario (the one with a SM-like heavier scalar $H$) and
in our sixth scenario (the one with \cp{} violation) we use the
charged-Higgs mass, $\mhp$, rather than $\mA$ as input parameter. In
that case we set the input flags of \FH\ as above, with the exceptions
of tlCplxApprox$\,=1$ in the fifth scenario and higgsmix$\,=3$,
tlCplxApprox$\,=3$ in the sixth scenario.

\bigskip

The branching ratios (BRs) for the decays of the neutral Higgs bosons
are also computed with \FH.\footnote{It should be noted that in this
  work we do not {strictly} follow the prescription of the
  \lhchxswg\ for the calculation of the
  BRs~\cite{Denner:2011mq}. However, the numerical differences are
  expected to be small, since they arise only from subdominant NLO
  corrections to some decay modes.} The decays to quark and lepton
pairs are evaluated at the full one-loop level, supplemented with
UV-finite wave-function normalization factors (the ``$Z$-factors'')
which ensure the correct on-shell properties of the external Higgs
bosons~\cite{Dabelstein:1995js, Heinemeyer:2001iy, Frank:2006yh,
  Williams:2007dc, Williams:2011bu, Fuchs:2016swt}, encoding the
effect of scalar mixing at the same loop level as in the Higgs-mass
calculation.  The decays to bottom pairs also include a resummation of
the $\tan\beta$-enhanced SUSY corrections~\cite{Banks:1987iu,
  Hall:1993gn, Hempfling:1993kv, Carena:1994bv, Carena:1999py,
  Carena:2000uj}, using one-loop formulae from~\citere{Hofer:2009xb}
for the so-called ``$\Delta_b$ terms''. The decays to gluons or
photons are computed at the lowest order (i.e., one loop),
supplemented with the NLO-QCD contributions from diagrams involving
gluons.  For the decays to massive gauge bosons, \FH\ approximates the
MSSM results by reweighting the SM results of the code {\tt
  Prophecy4F}~\cite{Bredenstein:2006rh, Bredenstein:2006ha} with the
appropriate Higgs{--}gauge{-}boson couplings. For the decays
to Higgs bosons \FH\ implements a full one-loop calculation within the
(complex) MSSM~{\cite{Williams:2007dc, Williams:2011bu}}, improved
with the resummation of potentially large logarithmic
corrections. Finally, the decays to SUSY particles are computed at the
tree level.

\subsection{Production cross sections}

\label{sec:production}

The cross sections for Higgs-boson production via gluon fusion and
bottom-quark annihilation are calculated with \sushi{
  1.7.0}~\cite{Harlander:2012pb, Harlander:2016hcx} and with its
extension to the MSSM with complex parameters, {\tt
  SusHiMi}~\cite{Liebler:2016ceh}. A link to \FH\ provides both the
loop-corrected Higgs-boson masses and the matrix of $Z$-factors (the
implementation of the latter in \sushi\ was discussed in
\citere{Liebler:2016ceh}).

For both the top- and bottom-quark contributions to gluon fusion,
\sushi\ includes the full next-to-leading order (\nlo)
results~\cite{Spira:1995rr, Harlander:2005rq}. In addition,
\sushi\ includes the next-to-NLO (\nnlo) top-quark contributions in
the heavy-quark effective theory~\cite{Harlander:2002wh,
  Anastasiou:2002yz, Ravindran:2003um, Harlander:2002vv,
  Anastasiou:2002wq} and even, for the \sm-like scalar only, the
next-to-NNLO (\nklo{3}) contributions, in a threshold
expansion~\cite{Anastasiou:2014lda, Anastasiou:2015yha,
  Anastasiou:2016cez} which has been shown to closely match the exact
result~\cite{Mistlberger:2018etf}. Compared with the results compiled
by the \lhchxswg\ for the gluon-fusion cross section in the
SM~\cite{deFlorian:2016spz}, \sushi\ omits only NNLO effects
suppressed by powers of the top mass.
In the MSSM with real parameters, scalar-quark (squark) and gluino
contributions to gluon fusion are taken into account at
\nlo\ following \citeres{Degrassi:2010eu, Degrassi:2011vq,
  Degrassi:2012vt}, which rely on an expansion in inverse powers of
the superparticle masses. In the MSSM with complex parameters these
\nlo\ contributions are interpolated, {while the leading-order
  contribution incorporates the full phase dependence,} see
\citere{Liebler:2016ceh}.
The $\tan\beta$-enhanced SUSY contributions to the
Higgs--bottom{-}quark couplings are resummed using the $\Delta_b$
terms provided by \FH. The two-loop electroweak corrections to scalar
production mediated by light quarks are included by reweighting the SM
results of \citeres{Aglietti:2004nj, Bonciani:2010ms} with the
appropriate $Z$-factors.  The central renormalization and
factorization scales are chosen to be $\muR=\muF=m_\phi/2$ (where
$\phi$ is the produced Higgs boson). For the parton distribution
functions (PDFs) we use the central set of {\verb=PDF4LHC15_nlo_mc=}
and {\verb=PDF4LHC15_nnlo_mc=}~\cite{Butterworth:2015oua} for the
\nlo\ and the \nnlo/\nklo{3} contributions, respectively.

For Higgs-boson production in bottom-quark annihilation we employ the
cross sections provided for the \sm\ Higgs boson as a function of its
mass by the \lhchxswg. Previously the \lhchxswg\ recommended
``Santander-matched'' cross sections~\cite{Harlander:2011aa}, which
combined results obtained in the five-flavor
scheme~\cite{Harlander:2003ai} and in the four-flavor
scheme~\cite{Dittmaier:2003ej, Dawson:2003kb, Wiesemann:2014ioa}.  In
the definition of the new benchmark scenarios we follow the most
recent recommendation~\cite{deFlorian:2016spz}, and use cross sections
that consistently match between the two schemes and were obtained in
the context of soft-collinear effective theory~\cite{Bonvini:2015pxa,
  Bonvini:2016fgf} (those coincide with the cross sections obtained in
the so-called ``fixed order plus next-to-leading log'' (FONLL)
approach~\cite{Forte:2015hba, Forte:2016sja}).  However, we only take
into account the part proportional to the {bottom Yukawa coupling}
and omit the interference with the top-quark contribution.  This is
well justified in the regions where bottom-quark annihilation is
experimentally relevant.  The cross sections for the \sm\ Higgs boson
are reweighted with effective Higgs--bottom{-}quark couplings,
using an effective mixing angle (or, in the case of \cp\ violation,
the matrix of $Z$-factors) in the scalar sector, and taking into
account the resummation of $\tan\beta$-enhanced SUSY contributions as
described earlier. In principle, the cross section for the production
of a \cp-odd scalar in bottom-quark annihilation differs from the one
of a \cp-even scalar, but this difference is negligible for
\cp-odd-scalar masses beyond $100$~GeV. Therefore, the SM cross
section is also used to obtain a reweighted cross section for the
\cp-odd scalar.

In our study we also take into account theoretical uncertainties for
the gluon-fusion and bottom-quark-annihilation cross sections.  For a
thorough discussion of all potential sources of uncertainty for these
two production channels in the MSSM we point the reader to
\citere{Bagnaschi:2014zla}.
Here, we follow a simplified approach: For the production of a
\cp-even scalar via gluon fusion in the \mssm, the relative
PDF+$\alpha_s$ uncertainties are assumed to coincide with those for the
production of a SM Higgs boson of the same mass, which can be
determined from the above-mentioned {\tt PDF4LHC15} sets (in
particular, the largest deviation from the result obtained with the
central PDF set is used as a symmetric relative uncertainty). For the
production of a \cp-odd scalar via gluon fusion we generate a separate
set of relative PDF+$\alpha_s$ uncertainties, assuming the field
content of a \thdm\ (in the scenario with \cp\ violation, however, we
apply the SM-inspired estimate to all three neutral scalars). The
second source of uncertainty that we take into account for gluon
fusion is the renormalization-scale dependence, which is estimated
using the analytic approach described in \citere{Harlander:2016hcx}.
For this purpose we take the minimal and maximal value of $100$
equidistant scale choices between $\muR=m_\phi/4$ and $\muR=m_\phi$,
and we use their difference as a symmetric uncertainty. The
factorization-scale dependence, on the other hand, is known to be
subdominant~\cite{Bagnaschi:2014zla} and is not further considered. We
finally add the renormalization-scale uncertainty and the
PDF+$\alpha_s$ uncertainty in quadrature.

For bottom-quark annihilation we use the absolute uncertainties
provided by the \lhchxswg\ for the \sm\ Higgs boson, as a function of
its mass and of the center-of-mass energy. Those include symmetric
renormalization- and factorization-scale uncertainties, symmetric
uncertainties related to the bottom-quark mass value and to the
bottom-quark matching scale, and asymmetric PDF+$\alpha_s$
uncertainties.  We add all downward (upward) {shifts} in
quadrature, and transform the result into a total relative downward
(upward) uncertainty.  This relative uncertainty is applied to the
production of all \mssm\ Higgs bosons, independently of the
\cp\ nature of the scalar under consideration.

The cross sections for Higgs production through vector-boson fusion,
Higgsstrahlung and associated production with top quarks are computed
with \FH, which reweights the SM predictions provided by the
\lhchxswg\ with the appropriate \mssm/\sm\ ratios of the couplings
involved.  Finally, the cross section for charged-Higgs production via
$gg \to tbH^\pm$ {and the corresponding uncertainty are} read from
a $(\mhp,\tb)$ grid for the type-II \thdm\ provided by the
\lhchxswg\ -- relying on the calculations of \citeres{Berger:2003sm,
  Dittmaier:2009np, Flechl:2014wfa, Degrande:2015vpa,
  Degrande:2016hyf} -- and then reweighted with the $\Delta_b$
corrections to the bottom Yukawa coupling provided by \FH.

\vfill
\newpage

\subsection{Interference effects in Higgs production and decay}
 \label{sec:theory_interference}
 
If two or more admixed Higgs bosons are nearly mass-degenerate and
their Breit-Wigner propagators overlap, large interference effects
occur in processes that involve these Higgs bosons in the $s$-channel.
Rather than calculating the full process $I\to \sum_a \,h_a\to F$,
involving the initial state~$I$, the final state~$F$ and the exchange
of all three of the Higgs mass eigenstates, {in}
\citeres{Fuchs:2014ola, Fuchs:2016swt, Fuchs:2017wkq} an approximation
{was developed} that combines the separate predictions for the
production and decay of each mass eigenstate $h_a$ with the respective
interference contributions:
\begin{align}
  \sigma\left(I\to {\textstyle \sum_a}\, h_a\to F\right)~\simeq~
{\textstyle \sum_{a}}~\sigma(I\to h_a)\,
  (1+\eta_{a}^{IF})\,\text{BR}(h_a\to F)\,.
\end{align}

The calculation of the interference factors~ $\eta_{a}^{IF}\equiv
\eta(I \to h_a \to F)$ is carried out at leading order only, however
it takes into account the radiatively corrected Higgs masses, their
total widths $\Gamma_{h_a}$ and the $Z$-factors (the latter affect the
internal Higgs-boson propagators).  The advantage of this procedure is
that higher-order corrections to the production and decay processes
can be taken into account separately.  This factorization is well
justified if the total widths of the involved Higgs bosons are not too
broad compared to the masses, and only neglects loop diagrams that
connect initial and final states.  For a more detailed explanation of
this approximation we point the reader to
\citeres{Fuchs:2014ola,Fuchs:2016swt, Fuchs:2017wkq}.

In the so-called ``decoupling limit''~\cite{Gunion:2002zf}, realized
in \cp-violating scenarios when $\mhp \gg \mZ$, the lightest scalar
$h_1$ hardly mixes with the two heavier scalars due to the large mass
splitting, and thus remains almost purely \cp-even. In contrast, $h_2$
and $h_3$ become approximately mass-degenerate and can reach a sizable
admixture, resulting in a large destructive interference effect in
processes involving $h_{2,3}$ in the $s$-channel. The interference
{occurs} as a consequence of the propagator structure that is also
reflected in the $Z$-matrix, and is therefore a general feature of the
decoupling limit in \cp-violating scenarios.  Focusing on the
$h_2\!-\!h_3$ interference, the interference factors are defined as
\begin{align}
 \eta_2^{IF}=\eta_3^{IF}
 ~\equiv~ \eta(I \to h_{2,3} \to F)
 ~=~\frac{\sigma_{{\rm coh}}}{\sigma_{{\rm incoh}}} - 1\,,
 \label{eq:eta}
\end{align}
where we distinguish the coherent cross section $\sigma_{{\rm
    coh}}=\sigma(|h_2+h_3|^2)$ that sums up the amplitudes involving
$h_2$ and $h_3$ from the incoherent cross section $\sigma_{{\rm
    incoh}}=\sigma(|h_2|^2)+\sigma(|h_3|^2)$.  We implemented the
calculation of the interference factors in \sushi{} for the initial
states $I \in \left\lbrace gg,~b\bar b\right\rbrace$ and the final
states $F \in \left\lbrace\tau\tau,~b\bar b,~t\bar t\right\rbrace$.
The Higgs{-}boson propagators are numerically integrated for the
invariant mass of the final state, $m^F$, within $m^F_{{\rm min}, {\rm
    max}}=(m_{h_2}+m_{h_3})/2 \mp 5(\Gamma_{h_2}+\Gamma_{h_3})/2$, and
\sushi\ {provides} the results in its output file for each
combination of initial state, final state and intermediate Higgs
boson. If such factors are eventually stored in the {\tt ROOT} files
of the \lhchxswg, they will be directly available to the experimental
collaborations.

In the \cp-violating benchmark scenario that we will define in
section~\ref{sec:CPV}, by far the most relevant interference is
obtained for the process $b\bar b\to h_{2,3}\to \tau^+\tau^-$,
inducing a distortion of the exclusion bounds from searches of heavy
Higgs bosons in the $\tau^+\tau^-$ final state with respect to the
\cp-conserving case.


\section{Definition of the benchmark scenarios}
\label{sec:benchmarks}

In this section we propose six new benchmark scenarios for
\mssm\ Higgs searches at the \lhc.  All scenarios include a scalar
with mass around $125$~GeV and SM-like properties {over large
  parts} of the defined parameter space, and are compatible with
recent searches for superparticles. In each scenario we leave two free
parameters, such that searches for additional Higgs bosons can be
presented in two-dimensional planes: one of the free parameters is
always $\tan\beta$, while the other is either $\mA$ (in the first four
scenarios) or $\mhp$ (in the last two).  In all scenarios a
significant region of the considered plane is still allowed by the
searches for additional Higgs bosons at the \lhc.  In scans over the
$(\mA\,,\tan\beta)$ plane, $\tan\beta$ is varied between $0.5$ and
$60$ and $\mA$ between $70$~GeV and $2$~TeV, except in the fourth
scenario where we {focus} on the region with $1\leq\tb\leq20$ and
$100~{\rm GeV}\leq\mA\leq1$~TeV.

{Indirect constraints on the MSSM parameter space such as those
  from requiring the correct Dark Matter density, from flavor
  observables or from the muon $g-2\,$, however interesting, depend to a
  large extent on parameters that are not crucial to Higgs-boson
  phenomenology. Following the spirit of previous benchmark
  proposals~\cite{Carena:1999xa, Carena:2002qg, Carena:2013ytb} we
  refrain from imposing additional constraints of this kind, with the
  obvious exception of the EDM constraints that are relevant to the
  definition of our scenario with \cp\ violation, see section
  \ref{sec:CPV}.}

\subsection{\sm\ input parameters}

We follow the recommendation of the \lhchxswg\ in
\citere{deFlorian:2016spz} and make use of the following \sm\ input
parameters:
\begin{eqnarray}
&m_t^{\text{pole}}=172.5~\text{GeV},\quad
\alpha_s(\mZ)=0.118,\quad
G_F=1.16637\cdot 10^{-5}~\text{GeV}^{-2},\nonumber\\
&m_b(m_b)=4.18~\text{GeV},\quad
\mZ=91.1876~\text{GeV},\quad
\mW=80.385~\text{GeV}\,.
\end{eqnarray}
The dependence of the Higgs-boson properties on other quark and lepton
masses is not very pronounced{,} and we stick to the default
values of \FH.  The value of the top-quark pole mass recommended by
the \lhchxswg\ is below the current world average\,\footnote{Note,
  however, that the relation between the quoted world-average mass of
  the top quark and its pole mass is subject to debate, see
  e.g.~\citeres{Hoang:2008xm, Butenschoen:2016lpz, Nason:2017cxd}.}
of $173.21\pm 0.51\pm 0.71$\,GeV~\cite{Olive:2016xmw}. We emphasize
that, in our scenarios, a change of $0.7$~GeV in the top-quark mass
induces a variation of about $0.5$--$0.6$~GeV in the \mssm\ prediction
for the mass of the \sm-like scalar. This significant parametric
uncertainty should be kept in mind when considering whether a point in
the \mssm\ parameter space is definitely ruled out by the Higgs-mass
prediction. Indeed, a $2\sigma$ variation of the input value for
top-quark mass could be accommodated by enlarging the interval of $\pm
3$~GeV that we allow for \mh\ in view of the theoretical uncertainty
of the MSSM prediction by a further $\pm 1.5$ GeV.

\subsection{SUSY input parameters}

In principle, the definition of an \mssm\ scenario would require
choices for about a hundred parameters in the
soft{-}\susy-breaking Lagrangian. However, since we are interested
in the MSSM Higgs-boson phenomenology, in this study we focus on
scenarios in which the soft{-}\susy-breaking terms do not include
new sources of flavor violation.  Therefore, the precise values of the
soft{-}\susy-breaking mass and interaction terms for the first-
and second-generation scalar fermions (sfermions) have only a limited
effect on the predictions for the Higgs masses and mixing. We thus
consider a common soft{-}\susy-breaking mass $M_{\tilde f}=2$~TeV
for the first- and second-generation sfermions, compatible with the
current exclusion bounds on squark masses by
\atlas~\cite{Aaboud:2017bac, Aaboud:2017vwy} and
\cms~\cite{Sirunyan:2017cwe, Sirunyan:2017kqq, Sirunyan:2018vjp}, and
we set the corresponding Higgs{--}sfermion interaction terms $A_f$
to zero. The remaining soft{-}\susy-breaking parameters that
define our different scenarios are: the third-generation squark mass
parameters $M_{Q_3}$, $M_{U_3}$ and $M_{D_3}$; the third-generation
scalar-lepton (slepton) mass parameters $M_{L_3}$ and $M_{E_3}$; the
third-generation Higgs{--}sfermion interaction terms $A_t$, $A_b$
and $A_\tau$; the gaugino masses $M_1$, $M_2$ and $M_3$.  In the first
three of our scenarios we will not fix an input value for $A_t$, but
rather for the combination $X_t= A_t-\mu\cot\beta$ which enters the
left{--}right mixing term in the stop mass matrix and determines
the correction to the mass of a \sm-like scalar (we recall that $\mu$
is the Higgs/higgsino superpotential mass, which we treat as an
additional input parameter). In the scenario where the heavier
\cp-even scalar $H$ is SM-like, several SUSY input parameters are
defined as a function of the free parameter $\mhp$, and thus vary over
the considered $(\mhp\,,\tan\beta)$ plane. We will consider all SUSY
input parameters to be real, except in the scenario where we study
\cp\ violation, in which we introduce a non-zero {phase} for
$A_t\,$.

The inclusion of radiative corrections in the predictions for the
Higgs masses and production cross section requires that we specify a
renormalization scheme for some of the SUSY input parameters. To be
compatible with the two-loop calculations implemented in \FH\ and in
\sushi\ (see \citeres{Heinemeyer:1998np, Degrassi:2001yf,
  Brignole:2001jy, Brignole:2002bz, Dedes:2003km, Heinemeyer:2007aq,
  Hollik:2014wea, Hollik:2014bua, Hollik:2015ema} and
\citeres{Degrassi:2010eu, Degrassi:2011vq, Degrassi:2012vt},
respectively, for the details) we will employ on-shell definitions
for the parameters $\mA$, $\mhp$, $M_{Q_3}$, $M_{U_3}$,
$M_{D_3}$, $A_t$ and $A_b$, whereas $\mu$ and $\tan\beta$ must be
interpreted as \drbar-renormalized parameters at a scale that \FH\
takes by default equal to the pole mass of the top quark.

\subsection{Experimental constraints on the MSSM Higgs sector}

The parameter space of our benchmark scenarios is already partially
constrained by the {limits obtained from} the searches for
additional Higgs bosons at the LHC, and by the requirement that one of
the neutral scalars match the properties of the observed Higgs
boson. We evaluate the former constraints with the code {\tt
  HiggsBounds}~\cite{Bechtle:2008jh,
  Bechtle:2011sb,Bechtle:2013wla,Bechtle:2015pma}, and the latter with
the code {\tt HiggsSignals}~\cite{Bechtle:2013xfa}.
We stress, however, that the regions that are indicated in our plots
below as disfavored by the properties of the observed Higgs boson and
by the existing limits from Higgs searches are not meant to be
regarded as ``inaccessible'', but should actually be explored in the
experimental analyses.  Our study of the existing constraints cannot
truly replace a dedicated analysis of the proposed benchmark scenarios
by ATLAS and CMS, which alone would be able to combine the results of
different searches taking into account all correlations.
In this section we briefly summarize the relevant features of the two
codes used in our study.

\subsubsection{Constraints from LHC searches for additional Higgs bosons}

The code {\tt HiggsBounds} tests each parameter point for
$95\%$~C.L.~exclusion from Higgs searches at the LHC (as well as LEP
and {the} Tevatron -- however, those turn out to be irrelevant
here). First, the code determines the most sensitive experimental
search available -- as judged by the expected limit -- for each
additional Higgs boson in the model. Then, only the selected channels
are applied to the model, i.e.~the predicted signal rate for the most
sensitive search of each additional Higgs boson is compared to the
observed upper limit, and in case the prediction exceeds the limit the
parameter point is regarded as excluded. For more details on the
procedure, we direct the reader to \citere{Bechtle:2015pma}.

Among the searches for {additional} neutral Higgs bosons that are
relevant in constraining our scenarios, the latest version, {\tt
  5.2.0beta}, of {\tt HiggsBounds} includes:
ATLAS~\cite{Aaboud:2017sjh} and CMS~\cite{Sirunyan:2018zut} searches
for heavy Higgs bosons decaying to $\tau^+\tau^-$ pairs using about
$36\,\mathrm{fb}^{-1}$ of Run-2 data, as well as the CMS results from
Run~1~\cite{CMS:2015mca}; Run-1 and Run-2 searches by
ATLAS~\cite{Aad:2015kna, Aaboud:2017rel} and
CMS~\cite{Khachatryan:2015cwa, Sirunyan:2018qlb} for a heavy scalar
decaying to a $Z$-boson pair; Run-1 searches by
ATLAS~\cite{Aad:2015xja} and Run-2 searches by
CMS~\cite{Sirunyan:2017djm, Sirunyan:2017guj} for a heavy scalar
decaying to a pair of $125$-GeV scalars; Run-1 searches by
ATLAS~\cite{Aad:2015bua} and CMS~\cite{Khachatryan:2017mnf} for the
$125$-GeV scalar decaying to a pair of lighter (pseudo)scalars; Run-1
searches by ATLAS~\cite{Aad:2015wra} and
CMS~\cite{Khachatryan:2015lba} for a heavy pseudoscalar decaying to a
Z boson and the 125-GeV scalar.
By comparing these results with the predictions of \sushi\ and
\FH\ for production cross sections and decay branching ratios of the
{additional} neutral Higgs bosons, {\tt HiggsBounds} reconstructs
the 95\% C.L.~exclusion contours in our scenarios.  In the MSSM these
constraints are typically stronger for large values of $\tan\beta$,
due to an enhancement of the production cross section of the heavier
Higgs bosons in bottom-quark annihilation (in that case the most
relevant searches are those for the decay to a $\tau^+\tau^-$ pair).

{\tt HiggsBounds} also contains the available constraints from the
searches for a charged Higgs boson by ATLAS and CMS. Most relevant in
our scenarios are the constraints on the production of a light charged
Higgs via a decaying top quark, $t\to H^+ b$, with subsequent decay
$H^+ \to \tau^+ \nu$~\cite{Khachatryan:2015qxa, CMS:2016szv,
  Aad:2014kga, Aaboud:2018gjj}, as well as top-quark associated
$H^\pm$ production, with subsequent decays to
$\tau\nu$~\cite{Khachatryan:2015qxa, CMS:2016szv, Aad:2014kga,
  Aaboud:2018gjj} or $tb$~\cite{Khachatryan:2015qxa, Aad:2015typ,
  ATLAS:2016qiq} pairs.

{In order to} estimate the theoretical uncertainty in our
determination of the excluded regions, we rely on the uncertainty
estimates described in section~\ref{sec:production} for the
gluon-fusion and bottom-quark annihilation cross sections. The most
conservative (i.e., weakest) determination of the exclusion region is
obtained by taking simultaneously the lowest values in the uncertainty
range for both production processes of each of the heavier Higgs
bosons, while the least conservative (i.e., strongest) determination
is obtained by taking simultaneously the highest values in the
uncertainty range.

\subsubsection{Constraints from the Higgs boson observed at the LHC}

We use the code {\tt HiggsSignals} to test the compatibility of our
scenarios with the observed Higgs signal, by comparing the predictions
of \sushi\ and \FH\ for the signal strengths for Higgs production and
decay -- defined as $\mu^{\scriptscriptstyle I} \equiv
\sigma^{\scriptscriptstyle I}/\sigma^{\scriptscriptstyle
  I}_{\scriptscriptstyle {\rm SM}}$ for each production process
$I\rightarrow h$ and $\mu^{\scriptscriptstyle F}\equiv {\rm
  BR}^{\scriptscriptstyle F}/{\rm BR}^{\scriptscriptstyle
  F}_{\scriptscriptstyle {\rm SM}}$ for each decay $h\rightarrow F$ --
against the measurements of the Higgs signal rates by ATLAS and CMS.
The latest version, {\tt 2.2.0beta}, of {\tt HiggsSignals} includes
the combined ATLAS and CMS results from Run~1 of the
LHC~\cite{Khachatryan:2016vau}, as well as the available
ATLAS~\cite{ATLAS:2016gld, ATLAS:2018gcr, Aaboud:2017xsd,
  Aaboud:2017vzb, Aaboud:2017jvq, Aaboud:2017rss, Aaboud:2018xdt} and
CMS~\cite{CMS:2017rli, Sirunyan:2017exp, Sirunyan:2017khh,
  Sirunyan:2017dgc, Sirunyan:2017elk, Sirunyan:2018shy,
  Sirunyan:2018ygk, Sirunyan:2018mvw, Sirunyan:2018egh,
  Sirunyan:2018hbu} results using about $36\,\mathrm{fb}^{-1}$ of
Run-2 data. We determine the region of the parameter space compatible
with these measurements by performing a log-likelihood ratio test
within the two-dimensional MSSM scenario under consideration.  In
particular, we look for the ``best-fit'' point -- i.e., the point
yielding the minimum of the total $\chi^2$ value of the signal
strengths, $\chi^2_{\text{min}}$ -- over the full parameter plane, and
then consider all points with $\Delta \chi^2 = \chi^2 -
\chi_{\text{min}}^2 \le 6.18$ to be allowed.  This corresponds to a
$2\sigma$ confidence level in the Gaussian limit.  For reference, a SM
Higgs boson with mass $125.09$ GeV results in a total $\chi^2$ value
of $\chi^2_\text{SM}/\mathrm{ndf} = 93.8/100$, where $\mathrm{ndf}$ is
the number of degrees of freedom. The latter is given by the number of
observables, which in our setup is $n_\text{obs}=100$, minus the
number of model parameters, $n_\text{par}$ (for the SM with fixed
Higgs mass we have $n_\text{par} = 0$, whereas for our two-dimensional
MSSM planes we have $n_\text{par} = 2$). We remark that, in the five
of our scenarios in which the observed Higgs boson is identified with
the lightest scalar, the best-fit point is located in the decoupling
region where the additional Higgs bosons are all very heavy, and it
provides essentially as good a fit to the measured Higgs rates as the
SM. Even in the scenario where the observed Higgs boson is the heavier
\cp-even scalar $H$, which is necessarily away from the decoupling
region, the best-fit point is less than two units of $\chi^2$ away
from $\chi^2_\text{SM}$.

A complication of our procedure is that, to account for the
theoretical uncertainty of the mass calculation in \FH, we allow the
prediction for the mass of the neutral scalar that we identify with
the observed Higgs boson to lie in an interval of $\pm 3$~GeV around
the value measured at the LHC.\footnote{We recall that the theoretical
  uncertainty of the prediction for the Higgs mass arises from missing
  higher-order terms, and does not lend itself to a statistical
  interpretation. Therefore, we treat the $\pm 3$~GeV interval as a
  hard cut, and we do not include the prediction for the Higgs mass in
  the determination of the $\chi^2$ value.}  The predictions for cross
sections and branching ratios show a non-negligible dependence on the
Higgs-boson mass in this interval, potentially biasing the comparison
with the LHC measurements. We will however assume that this mass
dependence is approximately the same in the MSSM and in the SM, such
that the predictions for the signal strengths are approximately
independent of the Higgs mass in the allowed $\pm 3$ GeV
interval. Consequently, in each point of the parameter space the
signal strengths computed by \sushi\ and \FH\ with the value of the
Higgs-boson mass predicted by \FH\ can be compared directly to those
obtained by ATLAS and CMS, in which the measured production and decay
rates are normalized to the state-of-the-art SM predictions provided
by the \lhchxswg~\cite{deFlorian:2016spz} for the measured value of
the Higgs-boson mass.


\subsection{\boldmath{\mhsc} scenario}

\label{sec:first scenario}

In our first benchmark scenario, denoted as the ``\mhsc\ scenario'',
all superparticles are chosen to be {so} heavy that production and
decays of the \mssm\ Higgs bosons are only mildly affected by their
presence. In particular, the loop-induced SUSY contributions to the
couplings of the lighter \cp-even scalar are small, and the heavy
Higgs bosons with masses up to $2$~TeV decay only to \sm\ particles.
Therefore, the phenomenology of this scenario at the \lhc\ resembles
that of a type-II \thdm{} with \mssm-inspired Higgs couplings. The
SUSY input parameters are fixed as
\begin{eqnarray}
&M_{Q_3}=M_{U_3}=M_{D_3}=1.5~\text{TeV},\quad
M_{L_3}=M_{E_3}=2~\text{TeV}, \nonumber\\[2mm]
&  \mu=1~\text{TeV},\,\quad
M_1=1~\text{TeV},\quad M_2=1~\text{TeV},
\quad M_3=2.5~\text{TeV}, \nonumber\\[2mm]
&X_t=2.8~\text{TeV},\quad A_b=A_\tau=A_t\,.
\label{eq:firstscenario}
\end{eqnarray}

The masses of the third-generation squarks and that of the gluino are
safely above the current bounds from direct searches at the \lhc\ (see
\citeres{Sirunyan:2017kqq, Sirunyan:2018vjp, Aaboud:2017nfd,
  Aaboud:2017ayj, Aaboud:2017aeu, Sirunyan:2017xse, Sirunyan:2017wif,
  Sirunyan:2017leh} for the stops,
\citeres{Aaboud:2017phn,Sirunyan:2017kqq, Sirunyan:2018vjp,
  Aaboud:2017dmy,Aaboud:2017wqg, Sirunyan:2017kiw} for the sbottoms
and \citeres{Aaboud:2017bac, Aaboud:2017vwy, Sirunyan:2017cwe,
  Sirunyan:2017kqq, Sirunyan:2018vjp, Aaboud:2017dmy} for the
gluino). To give an example, the stop pole masses are
$m_{\tilde{t}_1}=1340$~GeV and $m_{\tilde{t}_2}=1662$~GeV, and for
$\tb=20$ the sbottom pole masses are $m_{\tilde{b}_1}=1480$~GeV and
$m_{\tilde{b}_2}=1521$~GeV. The value chosen for $X_t$ is close to the
one for which the maximal value of $\mh$ is obtained.  Due to the
relatively large value of $\mu$, chosen to ensure that the higgsinos
are as heavy as the electroweak gauginos, the largest loop-induced
SUSY effects arise through the resummation of the $\tb$-enhanced
corrections to the Higgs{--}bottom{-}quark couplings: the
value of $\Delta_b$ is about $0.2$ for $\tb=20$, and about $0.6$ for
$\tb=60$. We recall that in the decoupling limit, realized when $\mA
\gg \mZ$, the effects of the $\Delta_b$ resummation cancel out
{for} the couplings of the lighter \cp-even scalar, which becomes
\sm-like. However, such non-negligible values of $\Delta_b$ do affect
the couplings to bottom pairs of the heavy Higgs bosons.

\begin{figure}[t]
\begin{center}
\includegraphics[width=0.75\textwidth]{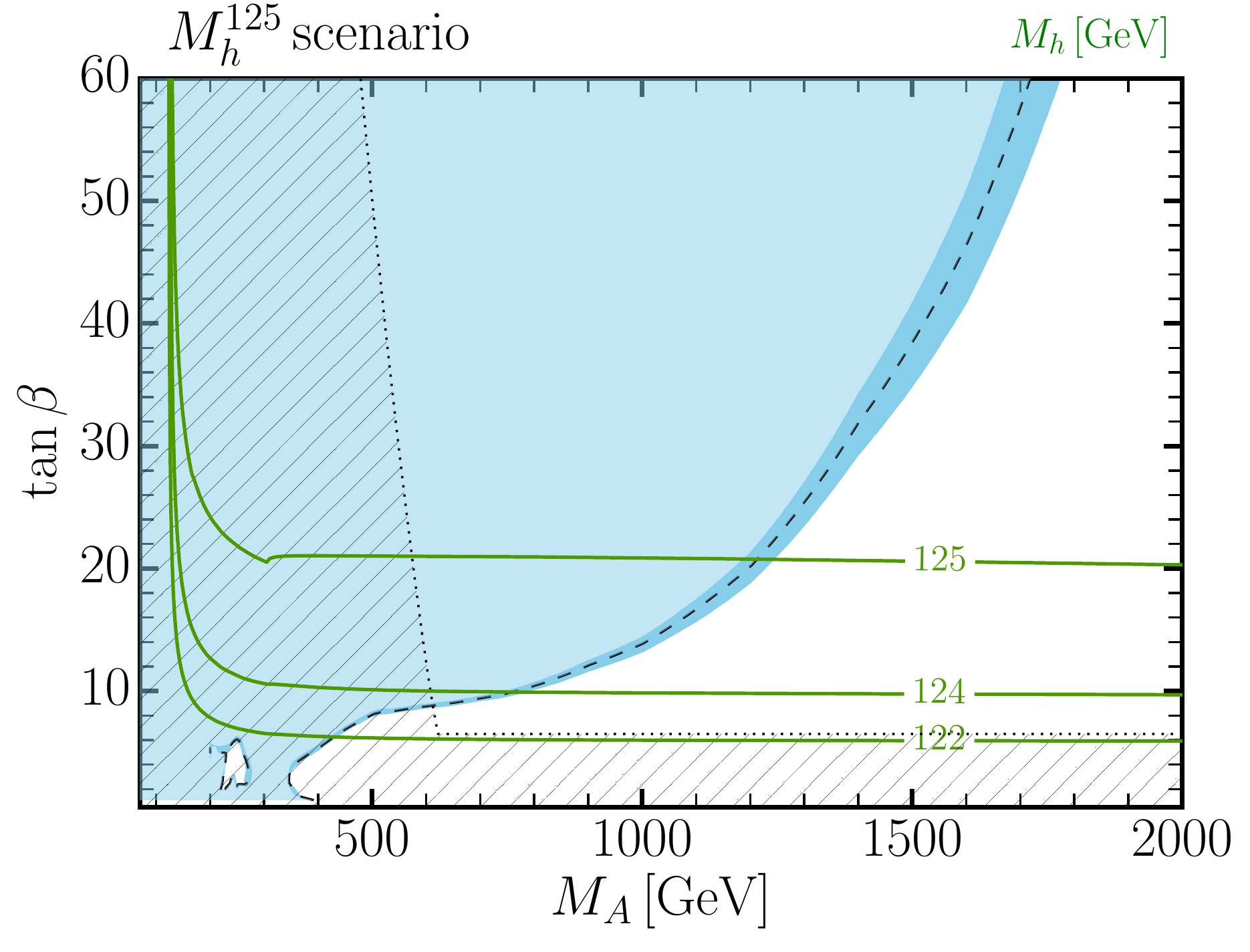} 
\end{center}
\vspace*{-5mm}
\caption{Constraints on the \mhsc\ scenario from Higgs searches at the
  LHC, in the $(\mA\,,\tb)$ plane. The green solid lines are
  predictions for the mass of the lighter \cp-even scalar $h$, the
  hatched area is excluded by a mismatch between the properties of $h$
  and those of the observed Higgs boson, and the blue area is excluded
  by the searches for additional Higgs bosons (the darker-blue band
  shows the theoretical uncertainty of the exclusion).}
\label{fig:mh125_mhiggs}
\end{figure}

In \fig{fig:mh125_mhiggs} we present, in the $(\mA\,,\tb)$ plane, the
existing constraints on the \mhsc\ scenario from Higgs-boson searches
at the \lhc. The blue area represents the bounds from searches for
heavy Higgs bosons ($H$ or $A$) as determined by {\tt HiggsBounds},
with the darker-blue band indicating the theoretical uncertainty of
the exclusion.  The green solid lines are the predictions of \FH\ for
the mass of the lighter \cp-even scalar $h$: we see that in this
scenario -- for the top-mass value recommended by the \lhchxswg\ --
$\mh$ is always below $126$~GeV. Finally, the hatched area is ruled
out because the light scalar cannot be identified with the Higgs boson
observed at the \lhc: for low $\tb$ the MSSM prediction for $\mh$
falls outside the window $125.09 \pm 3$~GeV, and for low $\mA$ the
MSSM predictions for the production and decay rates of $h$ are
determined by {\tt HiggsSignals} to be inconsistent with the
\lhc\ results.
In particular, the almost vertical exclusion boundary at {$\mA
  \approx 500\!-\!600$~GeV} reflects the behavior of the $hb\bar b$
coupling, which determines the {partial} width for the dominant
decay channel $h\rightarrow b\bar b$, and consequently affects the
branching ratios for all the subdominant decay channels. We recall
that, at the tree level, the $hb\bar b$ coupling is rescaled with
respect to its SM value by a factor of $-\sin\alpha/\cos\beta$, where
$\alpha$ is the mixing angle that rotates the neutral-scalar
components of $H_1$ and $H_2$ into the mass eigenstates $h$ and
$H$. For moderately large $\tb$, the tree-level rescaling factor is
well approximated by $1+2\,\mZ^2/\mA^2 + {\cal O}(\mZ^4/\mA^4)$,
i.e.~it enhances the coupling at moderate $\mA$, then tends to the SM
value in the decoupling limit of large $\mA$. The residual (mild)
$\tb$ dependence of the exclusion boundary at low $\mA$ is due to the
combined effects of the $\Delta_b$ corrections and of $\tb$-enhanced
stop-loop contributions to the Higgs mixing, see
\citere{Carena:2014nza}, which in this scenario suppress the $hb\bar
b$ coupling at large $\tb$.

\fig{fig:mh125_mhiggs} shows that a significant region of the
$(\mA\,,\tb)$ plane of the \mhsc\ scenario is still allowed by the
\lhc\ results from Higgs searches: {it starts at $\mA\approx
  600$~GeV and $\tb \approx 6\!-\!8$}, and it opens up to higher
values of $\tb$ for increasing $\mA$. The constraints at high values
of $\tb$ {arise} essentially from the searches for $H/A\rightarrow
\tau^+\tau^-$ at the \lhc\ with $13$~TeV center-of-mass
energy~\cite{Aaboud:2017sjh,Sirunyan:2018zut}.  On the other hand,
{values of $\tb$ lower than about $6$} are ruled out in the
\mhsc\ scenario by the prediction of a mass below $122.09$~GeV for the
\sm-like scalar.
The hole in the blue area {around $\mA\approx 250$~GeV and $\tb
  \approx 4$} corresponds to a region of the parameter space where $H$
has significant branching fractions to $ZZ$ and $hh$ pairs, but no
individual search is strong enough to yield an exclusion. However,
this region is ruled out by the requirement that the properties of $h$
match those of the observed Higgs boson.

\subsection{Scenarios with light superparticles}
\label{sec:lightSUSY}

Light superparticles, in particular charginos and neutralinos -- which
we collectively denote as electroweak (\EW)-inos -- and
third-generation sfermions, can substantially influence the Higgs
phenomenology, {see e.g.~\citeres{Carena:2011aa, Benbrik:2012rm,
    Carena:2013ytb, Carena:2013iba, Batell:2013bka, Belyaev:2013rza,
    Liebler:2015ddv}}.  This may happen through loop contributions to
the Higgs boson couplings to \sm\ particles, as well as, when
kinematically possible, through direct decays of the Higgs bosons into
superparticles.

\bigskip

\citere{Carena:2013ytb} proposed a scenario with degenerate
soft{-}\susy-breaking parameters $M_{Q_3}=M_{U_3}= 500$~GeV and
stop mixing $X_t=1$~TeV, resulting in lighter and heavier stop masses
of about $325$~GeV and $670$~GeV, respectively. The parameters $M_1$,
$M_2$ and $\mu$ {were chosen} in such a way that the lighter stop
decays almost entirely to the lightest neutralino and a charm quark,
thus evading the Run-1 searches for stops at the \lhc. However, as
mentioned in section~\ref{sec:introduction}, such ``light-stop''
scenarios are now strongly constrained by monojet searches by
\atlas~\cite{Aaboud:2017phn} and CMS~\cite{Sirunyan:2017kiw}.
Moreover, in the light-stop scenario of \citere{Carena:2013ytb} the
recent refinements in the Higgs{-}mass calculation of \FH\ result
in a prediction for $\mh$ that undershoots the observed value by more
than $3$~GeV. Even if it was possible to evade the Run-2 stop searches
by further {tuning} the \EW-ino masses, a scenario with a
lighter-stop mass of about $300\,$--$\,400$~GeV would require a large
splitting between $M_{Q_3}$ and $M_{U_3}$ in order to raise the
prediction for $\mh$ through radiative corrections involving the
heavier stop. However, in that case the accurate resummation of large
logarithmic effects would require a {specific} EFT setup -- not
yet implemented in \FH\ or in any other public code -- in which some
of the squark masses are close to the \EW\ scale while others are in
the multi-TeV region, see e.g.~\citeres{Espinosa:2001mm,
  Carena:2008rt}. Moreover, the results of \citeres{Degrassi:2010eu,
  Degrassi:2011vq, Degrassi:2012vt}, used by \sushi\ to compute the
two-loop SUSY contributions to the gluon-fusion cross section, rely on
a heavy-SUSY expansion valid only when the mass of the produced Higgs
boson is less than twice the lighter-stop mass, leading to an upper
bound on the values of $\mA$ allowed in our analysis. In view of these
limitations in the codes, and of the {tuning} of the MSSM
parameters that would be required to evade the bounds from squark
searches at the LHC, we refrain from proposing a new ``light-stop''
scenario for the time being.

In contrast, the bounds on light non-colored superparticles are still
weak, and can bring in phenomenologically interesting aspects.  For
certain observables, e.g.~the muon $g-2\,$, the contributions of
non-colored states can significantly reduce the tension between the
\sm\ prediction and experimental measurements, see
e.g.~\citeres{deVries:2015hva,Bechtle:2016kui}.
For what concerns Higgs phenomenology, the rate of the loop-induced
decay of the {lighter \cp-even scalar} to two photons can be
significantly altered by the contributions of light staus or light
charginos.  In particular, the former are enhanced for large values of
$\mu\tb$, while the latter are enhanced for low values of $\tb$
through the wino{--}higgsino mixing.  In addition, the decay of
the {lighter \cp-even scalar} to bino-like neutralinos can be
relevant if the latter are sufficiently light.  For the heavy Higgs
bosons, the decays to the light superparticles open up, and they
accordingly reduce the branching ratios into \sm\ particles.

In this paper we introduce two scenarios with light superparticles:
one with light staus and light EW gauginos, and another in which all
of the sfermions are heavy but all of the EW-inos (i.e., both gauginos
and higgsinos) are light. {They can be viewed as an update of the
  light-SUSY scenarios previously introduced in
  \citere{Carena:2013ytb}.}


\subsubsection{\boldmath{\lstau} scenario}
\label{section:lightstau}

In this scenario the SUSY input parameters are fixed as
\begin{eqnarray}
&M_{Q_3}=M_{U_3}=M_{D_3}=1.5~\text{TeV},\quad
M_{L_3}=M_{E_3}=350~\text{GeV}, \nonumber\\[2mm]
&  \mu=1~\text{TeV},\,\quad
M_1=180~\text{GeV},\quad M_2=300~\text{GeV},
\quad M_3=2.5~\text{TeV}, \nonumber\\[2mm]
&X_t=2.8~\text{TeV},\quad A_b=A_t,\quad A_\tau = 800~\text{GeV}\,.
\label{eq:stauscenario}
\end{eqnarray}

The parameters that determine the stop, sbottom and gluino masses are
the same as in the $\mh^{125}$ scenario, see \eqn{eq:firstscenario},
but the soft{-}SUSY-breaking masses and trilinear interaction term
for the staus are considerably reduced.  The left{--}right mixing
term in the stau mass matrix is $m_\tau X_\tau$, where $X_\tau =
A_\tau - \mu\tb$, thus the splitting between the two stau mass
eigenvalues increases with $\tb$.
The EW-gaugino masses $M_1$ and $M_2$ are in turn reduced with respect
to their values in the $\mh^{125}$ scenario.  Due to the hierarchy
among the parameters $M_1$, $M_2$ and $\mu$, the EW-ino spectrum is
{essentially} not mixed: the lightest neutralino is mostly bino
with mass around $180$~GeV, the lighter chargino and the
second-lightest neutralino are mostly winos with masses around
$300$~GeV, and the heavier EW-inos are mostly higgsinos with masses
around $1$~TeV.
Our choices of parameters in the stau and EW-ino sectors ensure that
the lightest SUSY particle (LSP) is the lightest neutralino, except
for the region with $\tb \gtrsim 52$ where the large value of $X_\tau$
{causes} the lighter stau $\tilde \tau_1$ to become the
LSP. However, as we discuss below, such high values of $\tb$ are
almost entirely excluded by other constraints.

The combined lower bounds on the lighter-stau and lighter-chargino
masses from the \lep\ experiments are $m_{\tilde{\tau}_1}\gtrsim
90$\,GeV~\cite{LEPstaulimit} and $m_{\tilde{\chi}^\pm_1} >
103.5$\,GeV~\cite{LEPchilimit}, respectively, i.e.~well below the
values that characterize the \lstau\ scenario. At the Run 2 of the
LHC, searches for direct stau-pair production are not yet sensitive
enough to constrain this scenario~\cite{Sirunyan:2018vig}. For what
concerns the EW-inos, the most relevant channels at the LHC are
$\tilde\chi_1^+\,\tilde\chi^-_1$ and
$\tilde\chi^\pm_1\,\tilde\chi^0_2$ pair production. The reach of these
searches in the \lstau\ scenario depends on the dominant decay
mechanism of the wino-like EW-inos. For $\tb \gtrsim 20$
the mass of the lighter stau is less than $300$~GeV, and the wino-like
EW-inos decay to the lightest, bino-like neutralino mostly via an
intermediate on-shell stau, resulting in final states with tau leptons
and missing energy.
In contrast, for lower values of $\tb$ the lighter stau is heavier
than $300$~GeV, and the dominant decays of the wino-like EW-inos are
to the lightest neutralino plus a gauge boson (or, when allowed, a
Higgs boson -- see \citere{Bharucha:2013epa}), in which case the most
sensitive searches are for final states with light leptons (electrons
or muons) and missing energy.
We checked that the current results of the LHC searches for EW-ino
pair production, both with~\cite{Aaboud:2017nhr, Sirunyan:2018vig} and
without~\cite{Aaboud:2018jiw, Aaboud:2018sua, Sirunyan:2018ubx} tau
leptons in the final state, {cannot exclude} an MSSM scenario with
bino mass around $180$~GeV and wino mass around $300$~GeV. An eventual
tightening of the bounds from LHC searches could be compensated for by
raising the value of $M_1$, at the price of lowering the value of
$\tb$ for which the lighter stau becomes the LSP.

\begin{figure}[t]
\begin{center}
\includegraphics[width=0.75\textwidth]{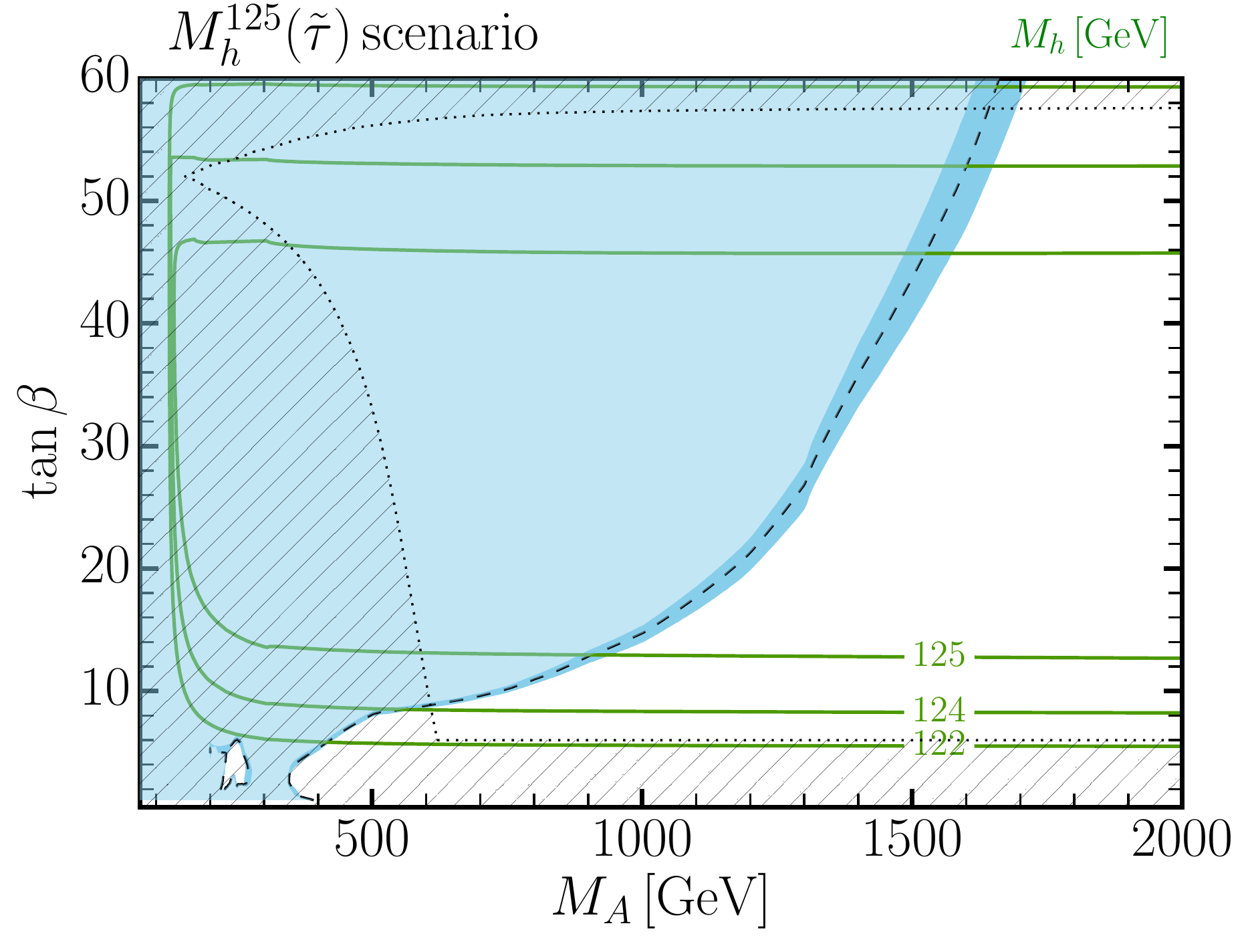}
\end{center}
\vspace*{-5mm}
\caption{Constraints on the \lstau\ scenario from Higgs searches at
  the LHC, in the $(\mA\,,\tb)$ plane. The green solid lines are
  predictions for the mass of the lighter \cp-even scalar $h$, the
  hatched area is excluded by a mismatch between the properties of $h$
  and those of the observed Higgs boson, and the blue area is excluded
  by the searches for additional Higgs bosons (the darker-blue band
  shows the theoretical uncertainty of the exclusion).}
\label{fig:mh125-ls_mhiggs}
\end{figure}

In \fig{fig:mh125-ls_mhiggs} we present, in the $(\mA\,,\tb)$ plane,
the existing constraints on the \lstau\ scenario from Higgs-boson
searches at the \lhc. The meaning of the different curves is the same
as in \fig{fig:mh125_mhiggs}. The blue region excluded by the LHC
searches for heavy Higgs bosons shows only a mild variation with
respect to the \mhsc\ scenario of \fig{fig:mh125_mhiggs}, namely a
small change in the slope of the exclusion boundary around $\mA
\approx 1.3$~TeV. On the other hand, significant changes are visible
in the hatched region excluded by the properties of the lighter
\cp-even scalar $h$. For moderate values of $\tb$, where the
corrections to the Higgs mass matrix that involve staus are not
important, the predictions of \FH\ for the mass of the lighter
\cp-even scalar (see the dashed lines) show a mild increase with
respect to the \mhsc\ scenario, which can be {traced back} to the
lower values adopted for $M_1$ and $M_2$ in the
\lstau\ scenario. Nevertheless, the prediction for $\mh$ remains below
$126$~GeV all over the $(\mA\,,\tb)$ plane, {and the lowest value
  of $\tb$ allowed by the $\pm3$~GeV theoretical uncertainty is about
  $6$}. In addition, for very large values of $\tb$ the corrections
involving staus cause the predictions for $\mh$ to decrease, reaching
the lowest allowed value of {$\mh = 122.09$~GeV for
  $\tb\approx60$}.
However, the hatched region in \fig{fig:mh125-ls_mhiggs} shows that
the requirement that the production and decay rates of the lighter
\cp-even scalar be SM-like rules out {values of $\tb$ larger than
  about $58$}.  This is in contrast with the \mhsc\ scenario, see
\fig{fig:mh125_mhiggs}, where at large $\tb$ the lighter \cp-even
scalar is sufficiently SM-like {as long as $\mA\gtrsim 500$~GeV}.
Finally, we remark that the region with $\tb \gtrsim 52$, in which the
lighter stau would be the LSP, is largely ruled out by the combination
of the blue and hatched regions.

\begin{figure}[p]
\begin{center}
\includegraphics[width=0.49\textwidth]{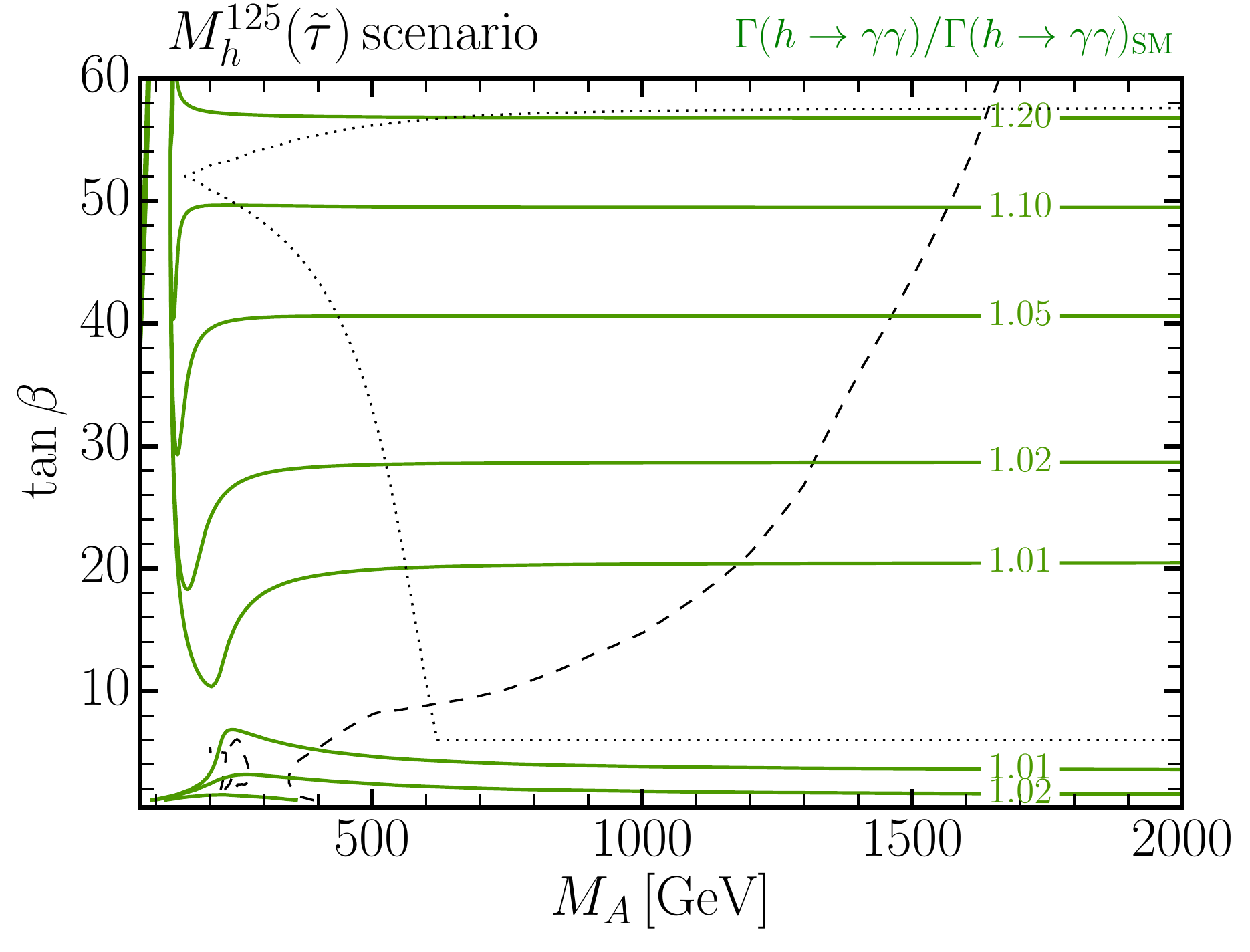}
\hfill
\includegraphics[width=0.49\textwidth]{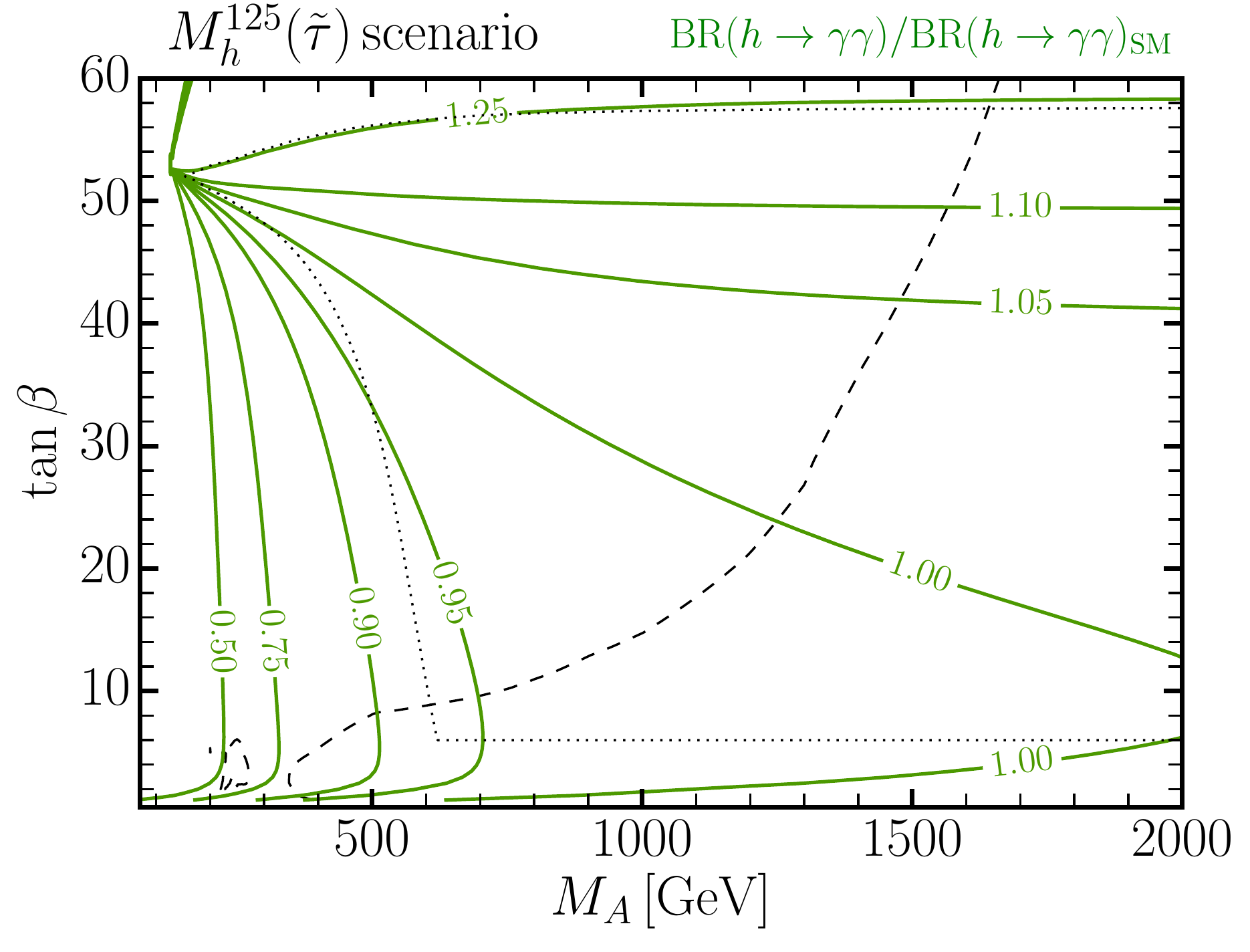}
\end{center}
\vspace*{-5mm}
\caption{Left: Decay width of the lighter \cp-even scalar into photons
  as a function of $\mA$ and $\tb$ in the \lstau\ scenario, normalized
  to the corresponding width of a \sm\ Higgs boson of the same
  mass. Right: same as the left plot for the branching ratio of the
  decay $h\rightarrow \gamma \gamma$. In each plot, the boundaries of
  the blue and the hatched exclusion regions of
  \fig{fig:mh125-ls_mhiggs} are also shown as a dashed and a dotted
  black line, respectively.}
\label{fig:mh125-ls_hgaga}
\end{figure}

\begin{figure}[p]
\begin{center}
\includegraphics[width=0.49\textwidth]{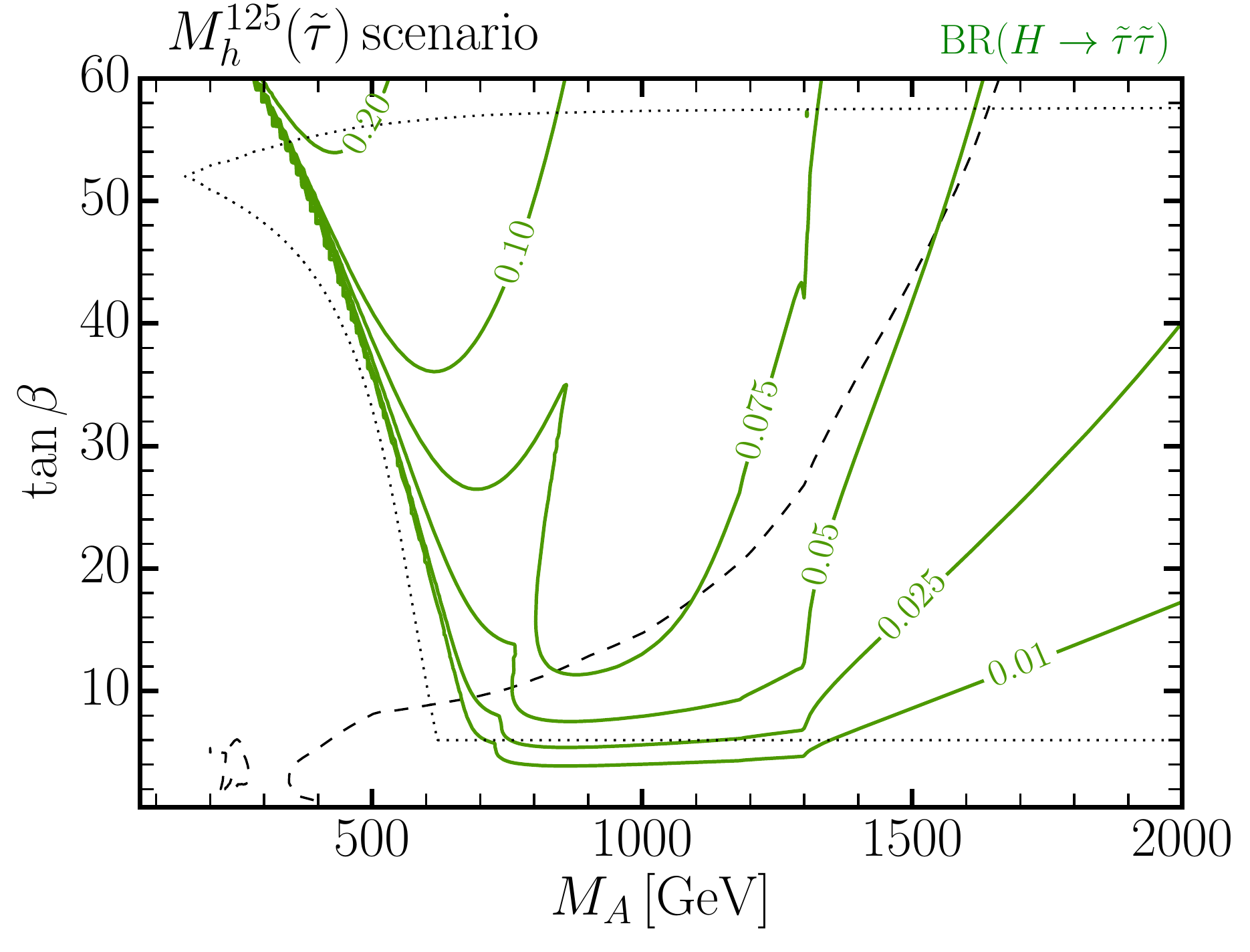}
\hfill
\includegraphics[width=0.49\textwidth]{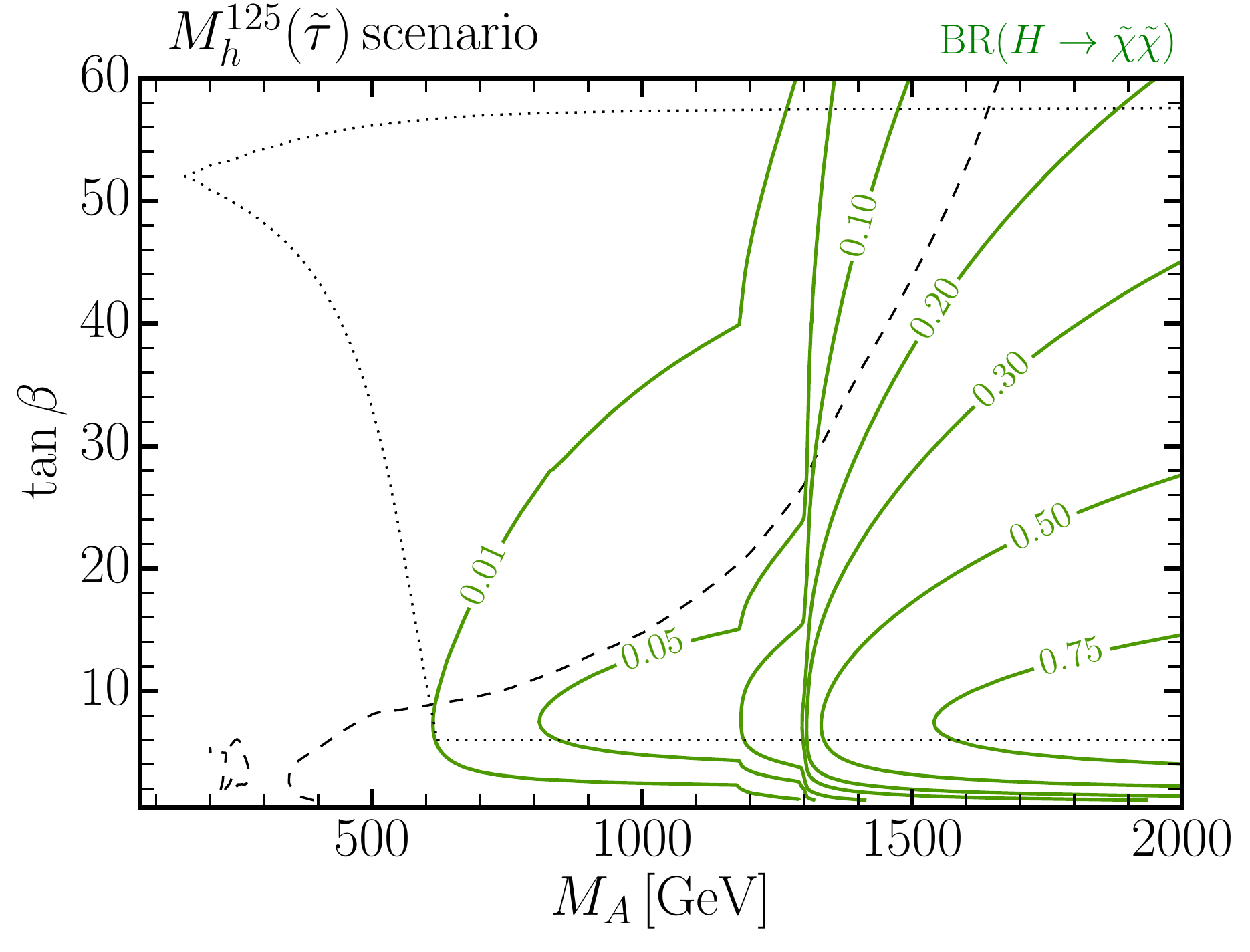}
\end{center}
\vspace*{-5mm}
\caption{Left: Total branching ratio for the decays of the heavier
  \cp-even scalar $H$ into stau pairs, as a function of $\mA$ and
  $\tb$ in the \lstau\ scenario. Right: same as the left plot for the
  total branching ratio of the decays of $H$ into chargino or
  neutralino pairs. In each plot, the boundaries of the blue and the
  hatched exclusion regions of \fig{fig:mh125-ls_mhiggs} are also
  shown as a dashed and a dotted black line, respectively.}
\label{fig:mh125-ls_HtoSUSY}
\end{figure}

In order to understand the shape of the exclusion regions in
\fig{fig:mh125-ls_mhiggs}, we now investigate how the presence of
light superparticles affects the decays of the MSSM Higgs bosons in
the \lstau\ scenario. As discussed, e.g., in \citere{Carena:2013iba},
both light-chargino and light-stau contributions can affect the
amplitude for the loop-induced decay of the lighter \cp-even scalar to
two photons, $h\rightarrow \gamma\gamma$. However, the chargino
contributions scale like $1/\tb$, and are therefore suppressed for the
moderate-to-large values of $\tb$ required to obtain $\mh \geq
122.09$~GeV. In contrast, the stau contributions contain a term
scaling like $\tan^2\beta$, and can therefore become relevant when
$\tb$ is sufficiently large. The green solid lines in the left plot of
\fig{fig:mh125-ls_hgaga} show, in the $(\mA\,,\tb)$ plane, the decay
width $\Gamma(h\rightarrow \gamma\gamma)$ as computed by \FH\ in the
\lstau\ scenario, normalized to the corresponding width of {a} SM
Higgs boson of the same mass. To guide the eye, the boundaries of the
blue and the hatched exclusion regions of \fig{fig:mh125-ls_mhiggs}
are also shown as a dashed and a dotted black line, respectively.  We
see that the EW-ino contributions {can exceed $1\%$ only in a
  low-$\tb$ region that is already excluded by the properties of the
  SM-like Higgs boson, as indicated by the dotted line,} whereas the
stau contributions can increase the diphoton width of the lighter
\cp-even scalar {by more than $10\%$ when $\tb\gtrsim50$}.

The right plot in \fig{fig:mh125-ls_hgaga} shows instead the branching
ratio for $h\rightarrow \gamma\gamma$, again normalized to the
corresponding quantity in the SM. We see that {also} the branching
ratio can become significantly enhanced at large $\tb$ -- resulting in
the excluded strip on the top margin of \fig{fig:mh125-ls_mhiggs} --
but the curves of constant MSSM/SM ratio show a markedly different
behavior from those in the left plot: indeed, when $\tb<50$ the
branching ratio can also be {\em suppressed} {by more than $10\%$,
  for values of $\mA$ extending up to about $500$~GeV}.
This is mainly due to the behavior of the $hb\bar b$ coupling, which
-- as described in section~\ref{sec:first scenario} -- is enhanced at
low $\mA$, thus suppressing the branching ratios for all subdominant
decay channels such as $h\rightarrow\gamma\gamma$, and then relaxes to
its SM value when $\mA$ is large enough.

For very large $\tb$, the corrections to the Higgs mixing angle due to
stau loops can also have a significant impact on the $hb\bar b$
coupling in the \lstau\ scenario. In particular, {for $\tb \approx
  52$} the scenario manifests the so-called ``alignment without
decoupling'' (see section \ref{sec:alignment} below), where the
$hb\bar b$ coupling remains SM-like for all values of $\mA$, so the
variation in the branching ratio for $h\rightarrow \gamma\gamma$ just
follows from the variation in the corresponding width. We note,
however, that for such large value of $\tb$ the LHC searches for $H/A
\rightarrow \tau^+\tau^-$ already rule out the region {with $\mA
\lesssim 1.6$~TeV}, see \fig{fig:mh125-ls_mhiggs}.

When kinematically allowed, the decays of the heavier \cp-even scalar
$H$ and of the \cp-odd scalar $A$ to light superparticles reduce the
branching ratios of their decays to tau pairs, potentially weakening
the exclusion bounds from the LHC searches. The left plot in
\fig{fig:mh125-ls_HtoSUSY} shows, in the $(\mA\,,\tb)$ plane, the
total branching ratio for the decays of $H$ to stau pairs, whereas the
right plot shows the total branching ratio for the decays of $H$ to
chargino or neutralino pairs. In both cases a sum is taken over all
the kinematically allowed combinations of particles in the final
state.  The plots show that the decays of $H$ to stau pairs can be
most relevant, {with a branching ratio above $20\%$}, in a region
with relatively small $\mA$ and large $\tb$ that is already well
excluded by the LHC searches for $H/A \rightarrow \tau^+\tau^-$. On
the other hand, the decays of $H$ to EW-inos can be most relevant for
$\mA\gtrsim 1.3$~TeV, when the final states involving a higgsino-like
EW-ino and a wino-like EW-ino are kinematically open. This explains
the mild change in the slope of the black dashed line (i.e., the
boundary of the region excluded by {\tt HiggsBounds}) around that
value of $\mA$. The right plot of \fig{fig:mh125-ls_HtoSUSY} also
shows that the total branching ratio for the decays of $H$ to EW-inos
is maximized, {and can exceed 75\%,} in the ``intermediate''
region {around $\tb \approx 7$}, where the decays of the heavy
Higgs bosons to $t\bar t$ pairs are {still} suppressed and those
to $b\bar b$ and $\tau^+\tau^-$ pairs are {only weakly} enhanced.
We remark that the decays of the \cp-odd scalar to EW-ino pairs follow
a similar pattern as the corresponding decays of the heavier \cp-even
scalar. However, for the decays of the \cp-odd scalar to staus only
the channel $A\rightarrow \tilde\tau_1\tilde\tau_2$ is open at tree
level, resulting in a maximal branching ratio {of about $14\%$} in
the (already excluded) region with {$\mA \approx 800$~GeV} and
large $\tb$.

Finally, it has long been known that charge- and color-breaking (CCB)
minima of the scalar potential can arise in the presence of large
left{--}right sfermion mixing~\cite{Nilles:1982dy, Frere:1983ag,
  AlvarezGaume:1983gj, Derendinger:1983bz, Claudson:1983et,
  Kounnas:1983td}.  In particular, vacuum stability bounds on the
parameters that affect the stau masses have been recently discussed in
\citeres{Hisano:2010re, Carena:2012mw, Kitahara:2013lfa}. The
requirement that the lifetime of the ordinary EW-breaking minimum be
longer than the age of the Universe constrains the combination
$\tb/(1+\Delta_\tau)$, where $\Delta_\tau$ represents non-decoupling,
$\tb$-enhanced SUSY corrections to the relation between the tau mass
and Yukawa coupling (they are analogous to the $\Delta_b$ effects in
the bottom/sbottom sector, but generally smaller as they involve only
the EW gauge couplings). For example, with the choices of SUSY
parameters given in \eqn{eq:stauscenario} the approximate bounds
presented in \citeres{Hisano:2010re, Kitahara:2013lfa} are violated
when {$\tb /(1+\Delta_\tau)\gtrsim 50$}. However, as discussed in
\citere{Carena:2012mw}, the vacuum stability bounds may also depend
non-trivially on the combination $A_\tau/(1+\Delta_\tau)$, as well as
on the \cp-odd scalar mass $\mA$. We have therefore used the code {\tt
  Vevacious}~\cite{Camargo-Molina:2013qva, Camargo-Molina:2014pwa,
  Wainwright:2011kj, Staub:2013tta} to check numerically the stability
of the ordinary EW-breaking minimum in the \lstau\ scenario all over
the $(\mA\,,\tb)$ plane, neglecting loop corrections to the potential
and thermal effects. We found that the ordinary EW-breaking minimum
can be unstable only in a narrow region at large $\tan\beta$ which is
already mostly excluded by the \lhc\ searches.  In a larger part of
the parameter space one or more deeper minima with non-zero stau vevs
exist, but the tunneling time from the ordinary minimum to
charge-breaking minima exceeds the lifetime of the Universe.


\subsubsection{\boldmath{\lchi} scenario}

In this scenario the SUSY input parameters are fixed as
\begin{eqnarray}
&M_{Q_3}=M_{U_3}=M_{D_3}=1.5~\text{TeV},\quad
M_{L_3}=M_{E_3}=2~\text{TeV}, \nonumber\\[2mm]
&  \mu=180~\text{GeV},\,\quad
M_1=160~\text{GeV},\quad M_2=180~\text{GeV},
\quad M_3=2.5~\text{TeV}, \nonumber\\[2mm]
&X_t=2.5~\text{TeV},\quad A_b=A_\tau=A_t\,.
\label{eq:lchiscenario}
\end{eqnarray}

The parameters $M_1$, $M_2$ and $\mu$ {have relatively small
  values that are} all close to each other, {giving rise to} a
significant mixing between higgsinos and gauginos and a compressed
EW-ino mass spectrum.  In particular, the mass of the lightest
neutralino $\tilde\chi^0_1$ varies between $105$~GeV for $\tb = 5$ and
$118$~GeV for the largest considered values of $\tb$, the mass of the
lighter chargino $\tilde\chi^\pm_1$ varies between $119$~GeV and
$131$~GeV in the same range of $\tb$, and the mass of the
second-lightest neutralino $\tilde\chi^0_2$ is about $165$~GeV
independently of $\tb$. The soft{-}SUSY-breaking masses for the
gluino and for the sfermions are the same as in the \mhsc\ scenario,
but the stop mixing parameter $X_t$ is slightly reduced, to partially
compensate for an enhancement in the prediction of the SM-like Higgs
mass due to light EW-inos. The sbottom mixing parameter $X_b=
A_b-\mu\tb$ is instead significantly {lowered} by the small value
of $\mu$.

Compressed EW-ino mass spectra are probed at the LHC by searches for
events with soft leptons and missing transverse momentum in the final
state~\cite{Sirunyan:2018iwl, Aaboud:2017leg}. In scenarios where the
slepton-mediated decays of the EW-inos are suppressed by large slepton
masses, the most sensitive channel is the production of a
$\tilde\chi^\pm_1\,\tilde\chi^0_2$ pair, followed by the decay of each
EW-ino into a virtual gauge boson -- which in turn decays to leptons
-- plus the lightest neutralino. However, the interpretation of the
LHC searches for EW-inos in this channel leads to the strongest bounds
when $\tilde\chi^\pm_1$ and $\tilde\chi^0_2$ are assumed to be
mass-degenerate pure winos. A full recast of those searches to the
\lchi\ scenario -- in which $\tilde\chi^\pm_1$ is a mixture of wino
and higgsino and $\tilde\chi^0_2$ is mostly bino and somewhat heavier
-- is beyond the scope of our paper and best left to the experimental
collaborations,
{but we did perform a naive study with {\tt CheckMATE}~\cite{Drees:2013wra,
    Dercks:2016npn,deFavereau:2013fsa,Cacciari:2011ma,Cacciari:2005hq,
  Cacciari:2008gp,Read:2002hq},
  using {\tt SDecay}~\cite{Muhlleitner:2003vg} to compute the decays
  of the EW-inos.  We found that the \lchi\ scenario is not
  constrained by the CMS and ATLAS searches of
  \citeres{Sirunyan:2018iwl, Aaboud:2017leg}, which were based on
  $36~{\rm fb}^{-1}$ of Run-2 data.\footnote{{Curiously, {\tt
        CheckMATE} does find constraints on the \lchi\ scenario
      arising from the search presented in a preliminary CMS note,
      \citere{CMS:2016zvj}, which was based only on $13~{\rm fb}^{-1}$
      of data. We could not find a convincing explanation for this
      apparent inconsistency.}} Some constraints appear to arise from
  a multilepton search by CMS~\cite{Sirunyan:2017lae}, but only at low
  values of $\tan\beta$ that are mostly ruled out by the prediction
  for $\mh$.}
  
\begin{figure}[t]
\begin{center}
\vspace*{-4mm}
\includegraphics[width=0.75\textwidth]{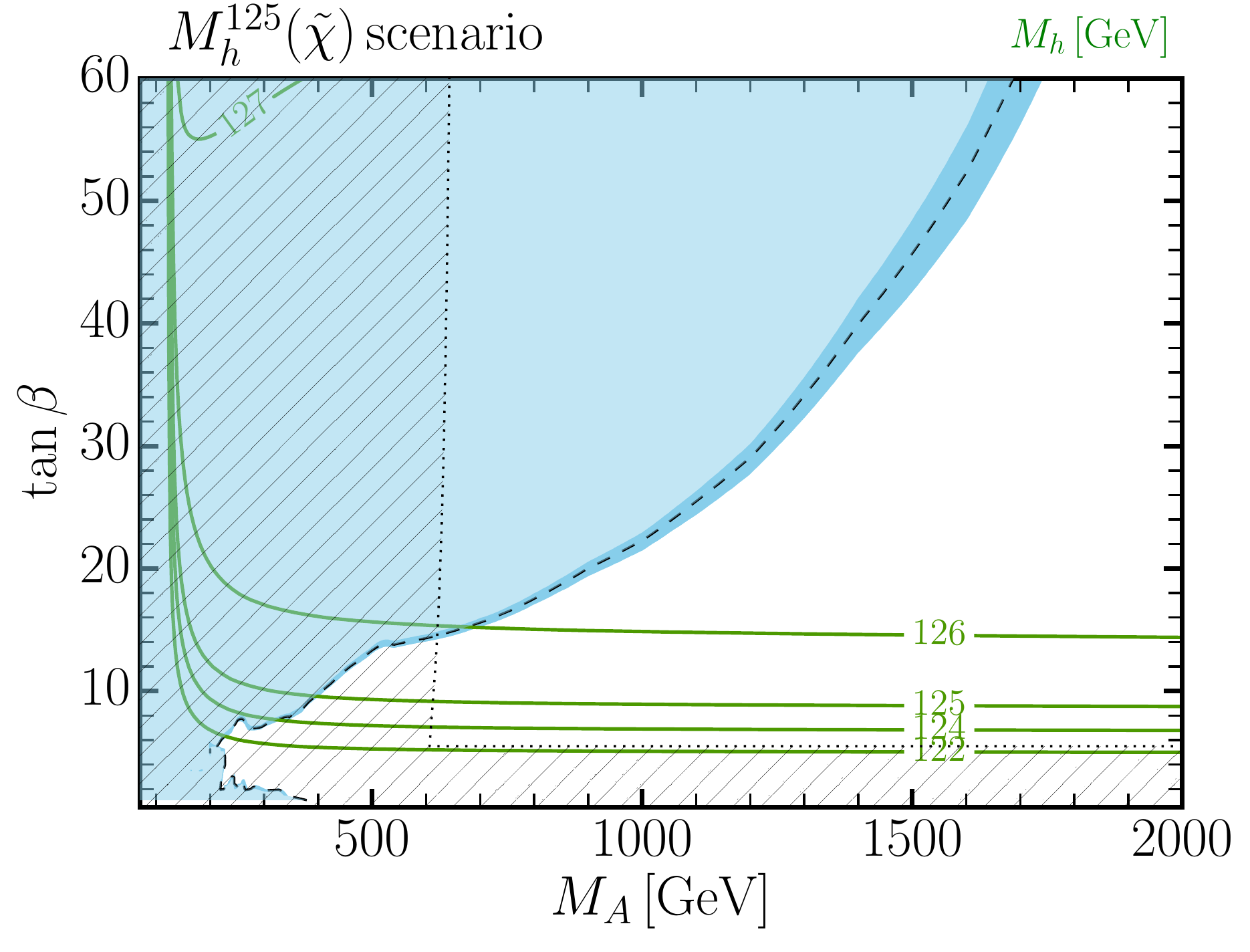}
\end{center}
\vspace*{-6mm}
\caption{Constraints on the \lchi\ scenario from Higgs searches at the
  LHC, in the $(\mA\,,\tb)$ plane. The green solid lines are
  predictions for the mass of the lighter \cp-even scalar $h$, the
  hatched area is excluded by a mismatch between the properties of $h$
  and those of the observed Higgs boson, and the blue area is excluded
  by the searches for additional Higgs bosons (the darker-blue band
  shows the theoretical uncertainty of the exclusion).}
\label{fig:mh125-lc_mhiggs}
\end{figure}

In \fig{fig:mh125-lc_mhiggs} we present, in the $(\mA\,,\tb)$ plane,
the existing constraints on the \lchi\ scenario from Higgs-boson
searches at the \lhc. The meaning of the different curves is the same
as in \fig{fig:mh125_mhiggs}. In the {lower}-left corner of the
plane, the blue region {that is} excluded by the LHC searches for
additional Higgs bosons {is significantly modified compared to}
the corresponding regions in the \mhsc\ and \lstau\ scenarios, see
\figs{fig:mh125_mhiggs} and \ref{fig:mh125-ls_mhiggs}. {This
  feature will be further discussed below.}
Concerning the hatched region excluded by the properties of the
lighter \cp-even scalar $h$, the reduced $\tb$ dependence (with
respect to \figs{fig:mh125_mhiggs} and \ref{fig:mh125-ls_mhiggs}) in
the boundary {around $\mA\approx 600$~GeV} is due to the small
value of $\mu$, which suppresses the $\tb$-enhanced corrections to the
$hb\bar b$ coupling.
Despite the reduction in $X_t$, the predictions for the mass of the
{lighter \cp-even} scalar show a mild increase with respect to the
\mhsc\ scenario.  However, they remain below $127$~GeV, except in the
upper-left corner of the $(\mA\,,\tb)$ plane. The lowest value of
$\tb$ allowed by the $\pm3$~GeV theoretical uncertainty in $\mh$ is
{about $5$}.

\begin{figure}[p]
\begin{center}
\includegraphics[width=0.49\textwidth]{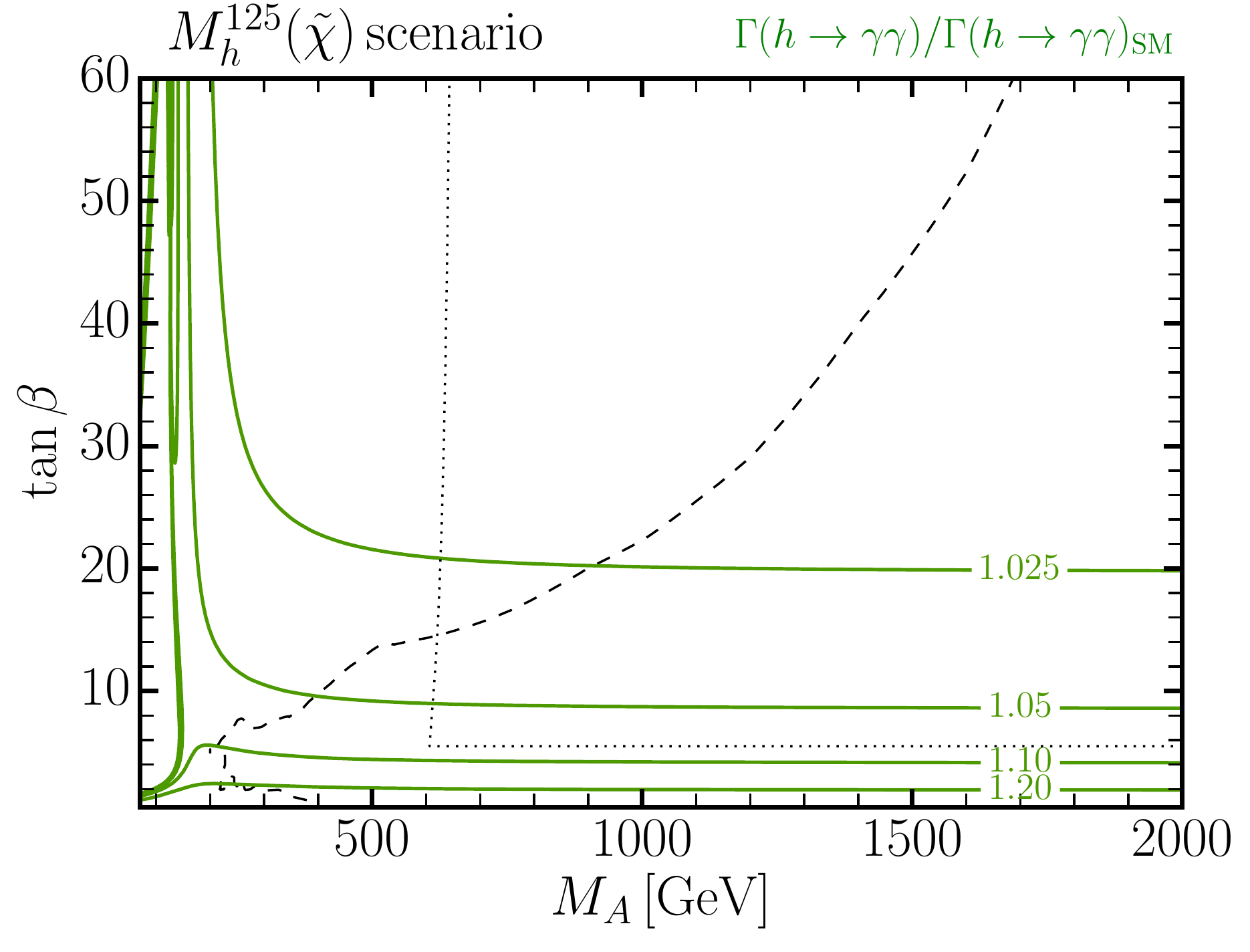}\hfill
\includegraphics[width=0.49\textwidth]{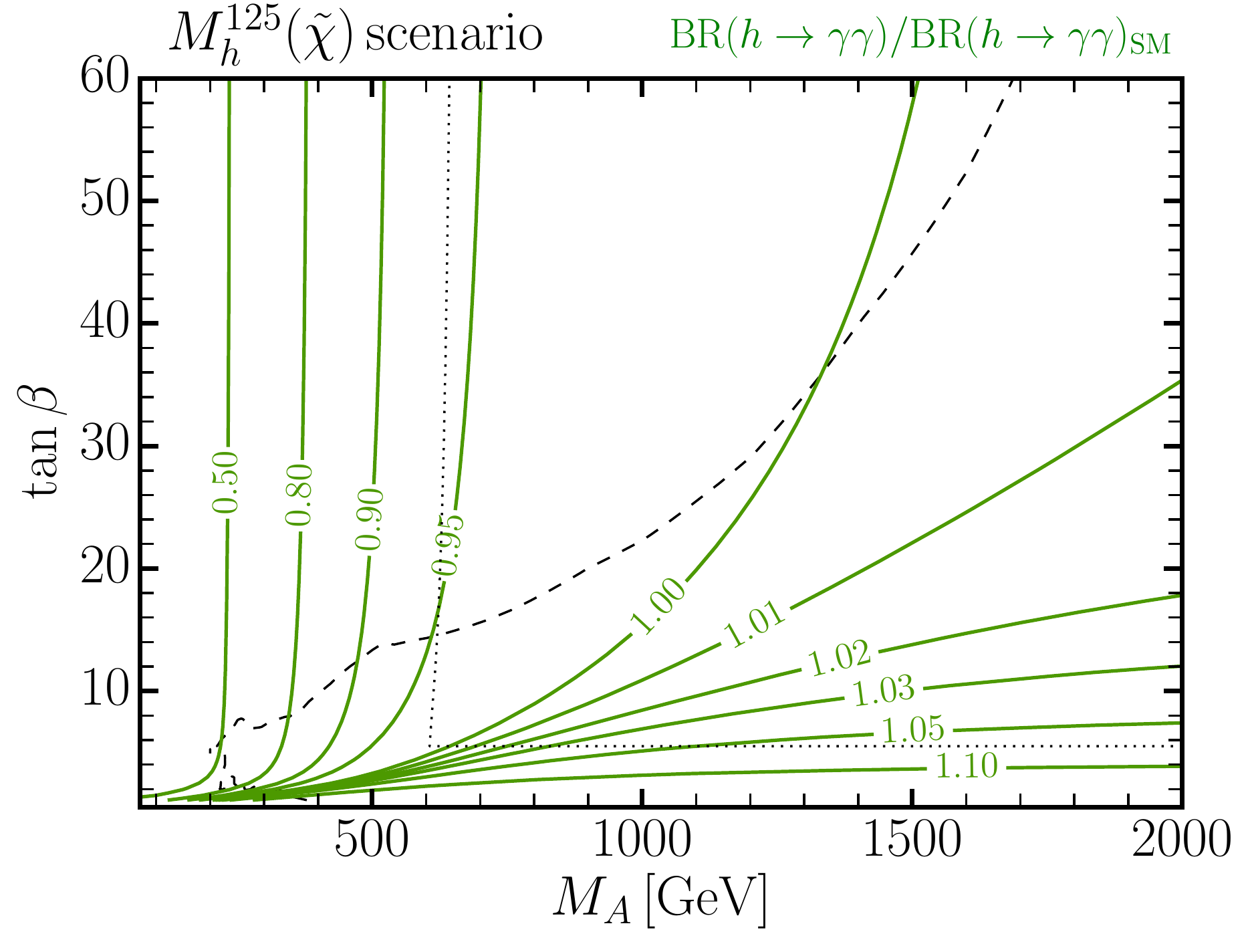}
\end{center}
\vspace*{-5mm}
\caption{Left: Decay width of the lighter \cp-even scalar into photons
  as a function of $\mA$ and $\tb$ in the \lchi\ scenario, normalized
  to the corresponding width of a \sm\ Higgs boson of the same
  mass. Right: same as the left plot for the branching ratio of the
  decay $h\rightarrow \gamma \gamma$. In each plot, the boundaries of
  the blue and the hatched exclusion regions of
  \fig{fig:mh125-lc_mhiggs} are also shown as a dashed and a dotted
  black line, respectively.}
\label{fig:mh125-lc_hgaga}
\end{figure}

\begin{figure}[p]
\begin{center}
\includegraphics[width=0.49\textwidth]{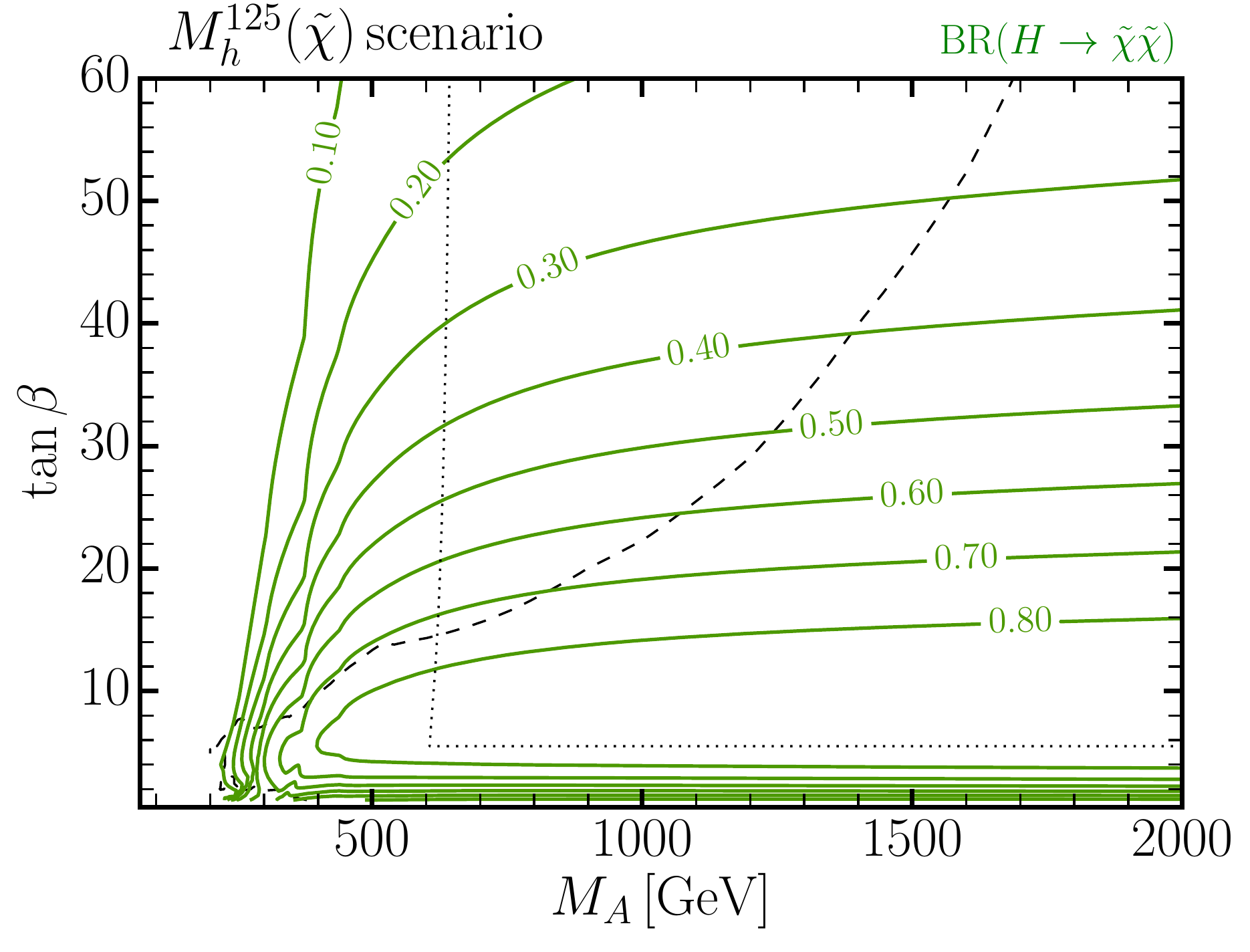}\hfill
\includegraphics[width=0.49\textwidth]{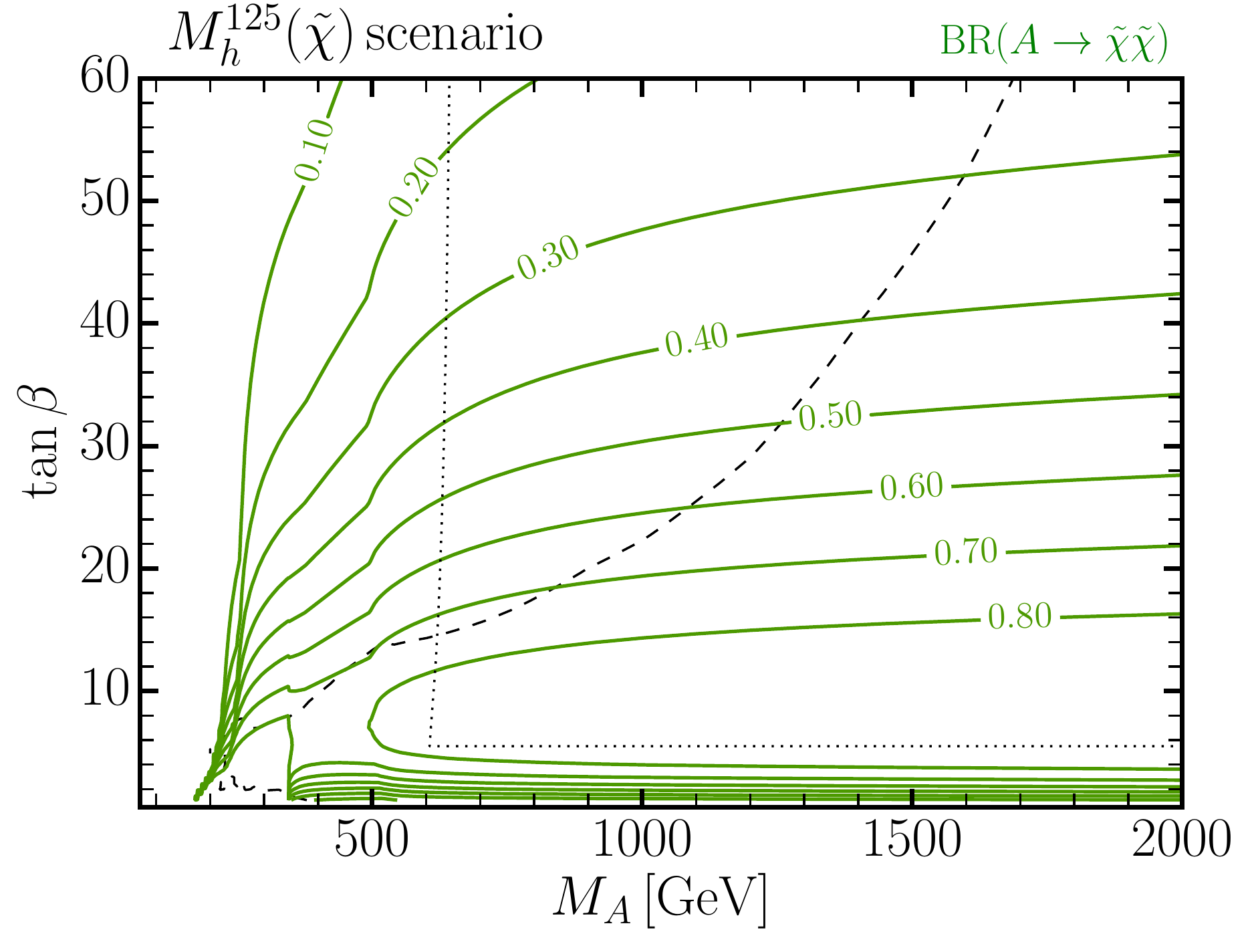}
\end{center}
\vspace*{-5mm}
\caption{Branching ratio for the decays of the heavier \cp-even scalar
  $H$ (left) or the \cp-odd scalar $A$ (right) into EW-ino pairs, as a
  function of $\mA$ and $\tb$ in the \lchi\ scenario. A sum is taken
  over all the kinematically allowed combinations of particles in the
  final state. In each plot, the boundaries of the blue and the
  hatched exclusion regions of \fig{fig:mh125-lc_mhiggs} are also
  shown as a dashed and a dotted black line, respectively.}
\label{fig:mh125-lc_HtoSUSY}
\end{figure}

We now discuss the effect of the presence of light EW-inos on the
decays of the MSSM Higgs bosons.  The left plot in
\fig{fig:mh125-lc_hgaga} shows, in the $(\mA\,,\tb)$ plane, the decay
width of the lighter \cp-even scalar to two photons in the
\lchi\ scenario, normalized to the corresponding width of a SM Higgs
boson of the same mass. As in the case of the \lstau\ scenario, see
the left plot in \fig{fig:mh125-ls_hgaga}, the EW-ino contributions
are most relevant at low values of $\tb$. However, the small value of
the higgsino mass parameter $\mu$ ensures that in the \lchi\ scenario
the effect on the prediction for $\Gamma(h\rightarrow \gamma\gamma)$
is much stronger, increasing it {by about $10\%$ for $\tb \approx
  4$}. Again, this effect must compete with the variation in the width
for the dominant decay channel $h\rightarrow b\bar b$, which is
enhanced at low $\mA$ and relaxes to its SM value at large $\mA$. This
is illustrated by the right plot in \fig{fig:mh125-lc_hgaga}, which
shows the branching ratio for the decay $h\rightarrow \gamma\gamma$ in
the \lchi\ scenario normalized to the corresponding quantity in the
SM. In addition to the enhancement of the EW-ino effects{, which
  can exceed $5\%$ in a sliver of the allowed region} at low $\tb$
{and large $\mA$}, the absence of stau-induced effects at large
$\tb$ explains the differences with the analogous plot for the
\lstau\ scenario, see \fig{fig:mh125-ls_hgaga}.

\fig{fig:mh125-lc_HtoSUSY} shows, in the $(\mA\,,\tb)$ plane, the
branching ratio for the decays of the heavier \cp-even scalar $H$
(left plot) or the \cp-odd scalar $A$ (right plot) to chargino or
neutralino pairs in the \lchi\ scenario. A sum is taken over all the
kinematically allowed combinations of particles in the final state,
and the dents visible in the curves for $\mA\lesssim 500$~GeV mark the
corresponding thresholds (as well as, at low $\tb$, the threshold for
the $t\bar t$ channel). The relatively low value of the higgsino mass
parameter $\mu$ in the \lchi\ scenario ensures that the decays to one
higgsino-like EW-ino and one wino-like EW-ino open up at much lower
values of $\mA$ than in the \lstau\ scenario. Indeed, for $5 \lesssim
\tb \lesssim 10$ the branching ratios for the decays of the heavy
Higgs bosons to EW-inos {can exceed $80\%$ already for $\mA\gtrsim
  500$~GeV}. In the region {around $\tb \approx 5$ and $\mA
  \approx 300$~GeV}, the dominance of the decays to EW-inos in the
\lchi\ scenario reduces the sensitivity of the $\tau^+\tau^-$ channel,
which explains the difference in the exclusion bounds from heavy-Higgs
searches with respect to the \mhsc\ and \lstau\ scenarios -- see
\figs{fig:mh125_mhiggs} and \ref{fig:mh125-ls_mhiggs}, respectively --
where this region is excluded by the LHC searches for $H/A \rightarrow
\tau^+\tau^-$.  However, we remark that this region is independently
ruled out by the requirement that the lighter \cp-even scalar be
sufficiently SM-like.
A comparison between the left and right plots of
\fig{fig:mh125-lc_HtoSUSY} shows that the decays of the \cp-odd scalar
follow patterns similar to those of the heavier \cp-even scalar in
most of the $(\mA\,,\tb)$ plane. The only exception is the corner with
low $\tb$ and $\mA$ below the $t\bar t$ threshold, where $H$ decays
mostly to pairs of SM particles whereas $A$ decays mostly to EW-ino
pairs.
The plots of \fig{fig:mh125-lc_HtoSUSY} suggest that a sizable region
of the parameter space of the \lchi\ scenario could be probed {by
  dedicated} searches for MSSM Higgs bosons decaying to EW-ino
pairs. {Very early studies of the discovery potential of those
  searches were presented in public notes by
  ATLAS~\cite{ATLAS:2009zmv}, which considered two scenarios from
  \citere{Bisset:2007mi}, and by CMS~\cite{Charlot:2006se,
    Ball:2007zza}. However,} we are not aware of any update taking
into account the existing results of the LHC runs.
{We also remark that an eventual tightening of the bounds on the
  EW-ino masses could be easily compensated for by an increase in the
  values of $M_1$, $M_2$ and $\mu$ by about $100$~GeV, without
  significantly affecting the decay rates of the heavy Higgs bosons in
  the allowed region with $\mA \gtrsim 600$~GeV. The main implication
  of such a modification of our scenario would be a smaller enhancement
  of the branching ratio for $h\rightarrow \gamma \gamma$ at low
  $\tb$.}

\begin{figure}[t]
\vspace*{-2mm}
\begin{center}
\includegraphics[width=0.49\textwidth]{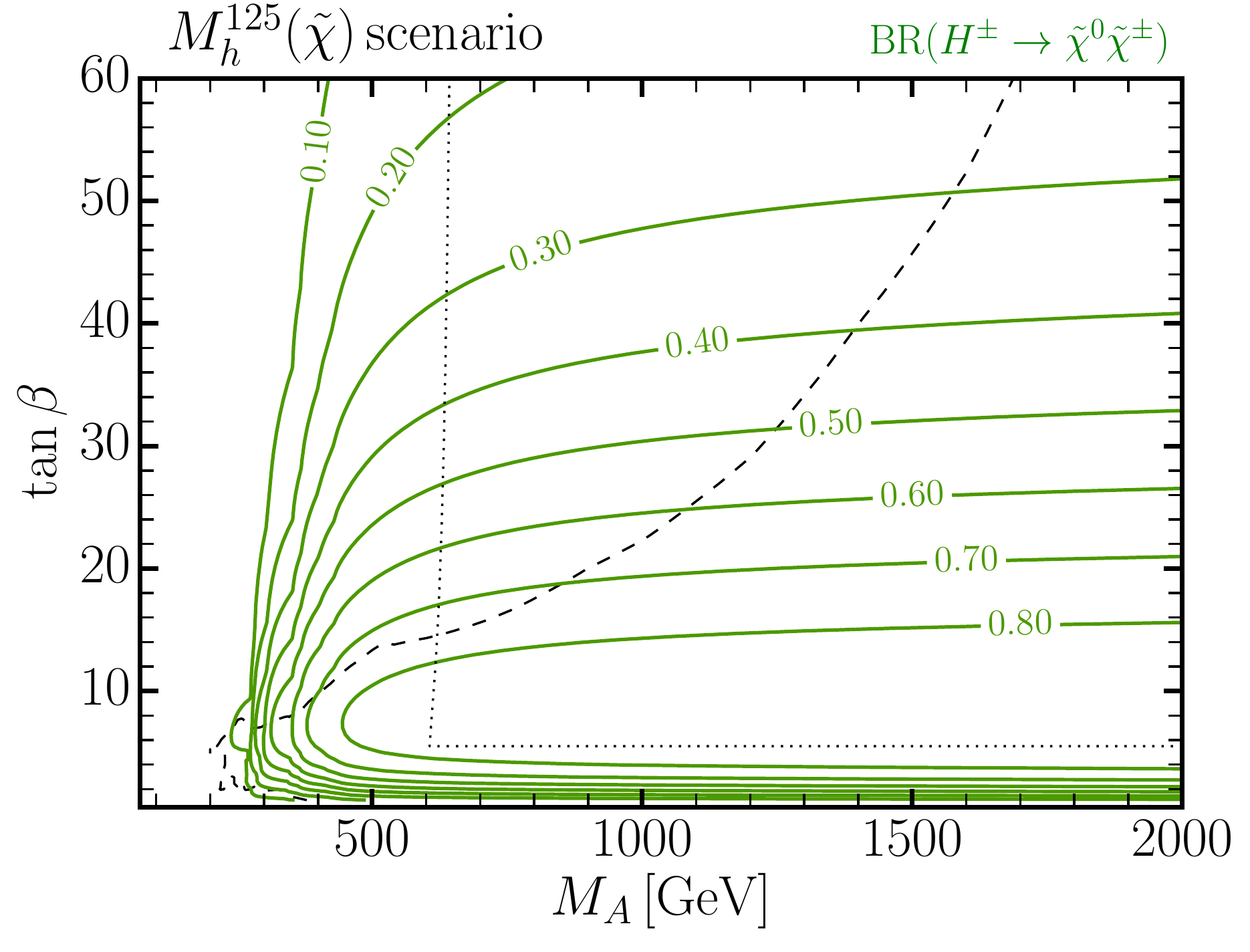}\hfill
\includegraphics[width=0.49\textwidth]{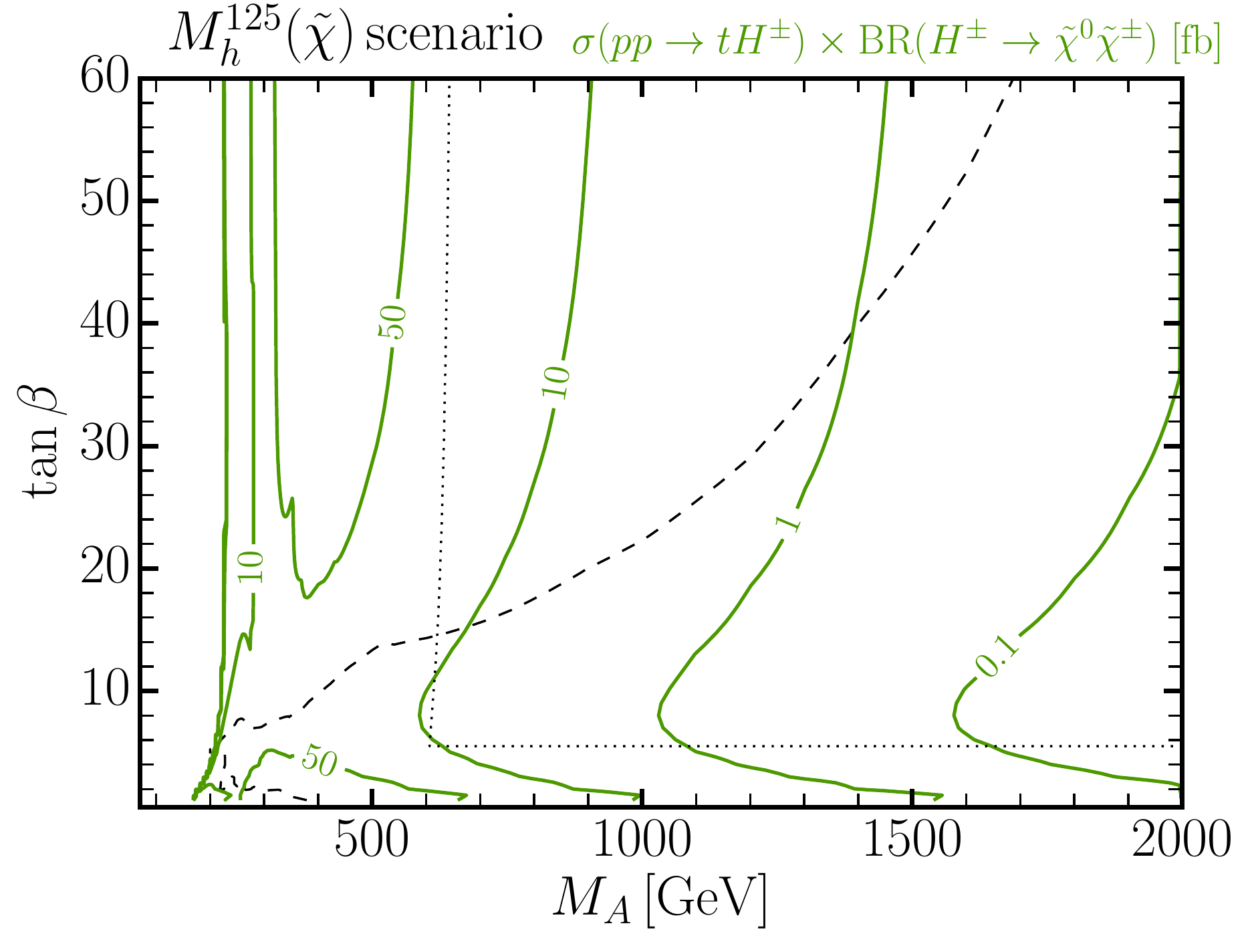}
\end{center}
\vspace*{-5mm}
\caption{Left: Branching ratio for the charged-Higgs decays to
  chargino{--}neutralino pairs, as a function of $\mA$ and $\tb$
  in the \lchi\ scenario. A sum is taken over all the kinematically
  allowed combinations of particles in the final state.  Right:
  {Total rate} (in fb) for {the} production of a charged Higgs
  boson in association with a top quark{, followed by its decay to
    EW-inos,} at the LHC with $13\,\mathrm{TeV}$ center-of-mass
  energy.  In each plot, the boundaries of the blue and the hatched
  exclusion regions of \fig{fig:mh125-lc_mhiggs} are shown as a dashed
  and a dotted black line, respectively.}
\label{fig:mh125-lc_Hpm}
\end{figure}

Finally, the left plot in \fig{fig:mh125-lc_Hpm} shows, in the
$(\mA\,,\tb)$ plane, the branching ratio for the decays of a charged
Higgs boson $H^\pm$ to chargino{--}neutralino pairs in the
\lchi\ scenario (again, a sum is taken over all of the allowed final
states). A comparison with \fig{fig:mh125-lc_HtoSUSY} shows that the
decays of $H^\pm$ to EW-inos follow approximately the same pattern as
the corresponding decays of the heavier \cp-even scalar, and can be
the dominant ones for intermediate values of $\tan\beta$.  The right
plot in \fig{fig:mh125-lc_Hpm} shows instead the {total rate} (in
fb) for the production of a charged Higgs boson in association with a
top quark{, followed by the decay to a chargino--neutralino pair,}
at the LHC with $13$ TeV center-of-mass energy {(a factor $2$ is
  included to account for both Higgs charges). The plot}
suggests that, for the values of the charged-Higgs mass allowed
{in this scenario} by the constraints on the neutral sector
{(say, $\mhp \gtrsim 600$~GeV)} and intermediate values of
$\tan\beta$, the rates for top-associated production of $H^\pm$
followed by decays to EW-inos are comparable to those recently probed
by ATLAS in the $\tau\nu$ channel with $36$~fb$^{-1}$ of Run-2
data~\cite{Aaboud:2018gjj}.
Even considering the complications in the reconstruction of the
subsequent EW-ino decays, we believe that a study of this scenario
with the full Run-2 dataset would be well motivated.


In summary, as was discussed e.g.~in
\citeres{Carena:2013ytb,Craig:2015jba, Barman:2016kgt,
  Profumo:2017ntc}, the presence of light EW-inos in the
\lchi\ scenario lowers the sensitivity of the traditional searches for
heavy Higgs bosons decaying to pairs of \sm\ particles, especially in
the region with intermediate $\tb$, but opens up the possibility to
probe the same region through the Higgs decays to SUSY particles. We
{propose these decays as a target} of dedicated searches at the
LHC. 


\subsection{Scenarios characterized by alignment without decoupling}
\label{sec:alignment}

In models with an extended Higgs sector, the so-called {\em alignment
  limit}~\cite{Gunion:2002zf,Craig:2013hca,Carena:2013ooa,
  Carena:2014nza,Bechtle:2016kui,Haber:2017erd} corresponds to the
case in which one of the neutral \cp-even scalars is aligned in field
space with the direction of the SM Higgs vev, and thus has SM-like
couplings to gauge bosons and matter fermions.
In a general \thdm, such alignment is commonly associated with the
{\em decoupling limit}, in which the lighter \cp-even scalar $h$ is
approximately SM-like and all the remaining Higgs states are
significantly heavier. However, alignment can also arise {\em without
  decoupling} -- i.e., irrespective of the Higgs mass spectrum -- if
the term in the mass matrix for the neutral \cp-even scalars that
mixes the field aligned with the SM Higgs vev with the field
orthogonal to it vanishes exactly. In this case, the role of the SM
Higgs boson can be played by either of the two \cp-even scalars.

Alignment without decoupling can arise as a consequence of some global
symmetry of the Higgs sector~\cite{Dev:2014yca, Pilaftsis:2016erj}, or
even of an extended supersymmetry~\cite{Antoniadis:2006uj,
  Ellis:2016gxa, Benakli:2018vqz}. In the \mssm, however, it can only
arise from an accidental cancellation between the tree-level mixing
term and the radiative corrections in the mass matrix, and is thus
associated to rather specific choices of the SUSY parameters. As
discussed in detail in \citeres{Carena:2013ooa,
  Carena:2014nza,Bechtle:2016kui,Haber:2017erd}, for values of
$\mu/M_S$ and $A_f/M_S$ of order one (where $A_f$ denotes collectively
the trilinear soft{-}SUSY-breaking couplings of the third-generation
sfermions, and $M_S$ denotes the scale of the corresponding sfermion
masses) the required cancellation in the mass matrix can be achieved
only for large values of $\tan\beta$. For example, we noted in
section~\ref{section:lightstau} that the \lstau\ scenario features
alignment without decoupling {for $\tan\beta\approx 52$}, thanks to
radiative corrections involving stau loops. However, already {for
  $\tan\beta\gtrsim 15$} the bounds from $\tau^+\tau^-$ searches at
the Run 2 of the LHC allow only for masses of the additional Higgs
bosons above one TeV, i.e.~large enough to be well within the
decoupling limit.  To obtain alignment without decoupling {for
  $\tan\beta\lesssim8$}, such that the $\tau^+\tau^-$ bounds allow for
masses of the additional Higgs bosons below $500$~GeV, the ratios
$\mu/M_S$ and, in some cases, $A_t/M_S$ must be increased to values of
about $3$ or even larger (in contrast, the couplings of staus and
sbottoms become less relevant for moderate $\tan\beta${, as do the
$\Delta_b$ effects\,\footnote{{Despite the large values of
  $\mu/M_S$, in the relevant regions of our alignment scenarios we
  find $\Delta_b \approx 0.2$\,.}}}). While we are aware that such large
values of $\mu/M_S$ and $A_t/M_S$ {are expected to} be in conflict
with the requirement {of avoiding CCB minima of the scalar
  potential}~\cite{Nilles:1982dy, Frere:1983ag, AlvarezGaume:1983gj,
  Derendinger:1983bz, Claudson:1983et, Kounnas:1983td}, we believe
that the rich Higgs-sector phenomenology of alignment scenarios with
moderate $\tan\beta$ still motivates their investigation at the LHC.

Benchmark scenarios for MSSM Higgs searches characterized by alignment
without decoupling have already been proposed in the literature, but
their viability has been investigated only with respect to the results
of Higgs and SUSY searches at Run~1 of the LHC. In particular,
\citeres{Carena:2014nza,Bechtle:2015pma} introduced a scenario named
``\,$m_h^{\rm alt}$\,'', in which alignment occurs for $\tb\approx 10$
and the role of the SM-like Higgs boson with mass around $125$~GeV is
played by $h$. \citere{Bechtle:2016kui} introduced instead three
variants of a scenario named ``\,low-$M_H^{\rm alt}$\,'' -- themselves
meant as updates of the original ``\,low-$M_H$'' scenario of
\citere{Carena:2013ytb} -- in which alignment occurs for $\tb\approx
6$--$7${,} and the role of the observed Higgs boson is played by the
heavier \cp-even scalar $H$ (the Dark Matter phenomenology of
alignment scenarios was subsequently investigated in
\citeres{Profumo:2016zxo,Duan:2017ucw}). An update of these alignment
scenarios seems now in order, motivated both by the improvements in
the {predictions for} Higgs masses and couplings implemented in
\FH\ and by the tightening of the experimental constraints on the
relevant MSSM parameters in view of the available Run-2 results.

In this section we present two new benchmark scenarios characterized
by alignment without decoupling at relatively low values of
$\tan\beta$. In the first scenario the role of the observed Higgs
boson is played by $h$, whereas in the second scenario -- which
requires somewhat extreme parameter choices and is already highly
constrained (but not ruled out) by the available LHC searches -- that
role is played by $H$.

\subsubsection{\boldmath{\halignment} scenario}

In this scenario the SUSY input parameters are
fixed as
\begin{eqnarray}
&M_{Q_3}=M_{U_3}=M_{D_3}=2.5~\text{TeV},\quad
M_{L_3}=M_{E_3}=2~\text{TeV}, \nonumber\\[2mm]
&  \mu=7.5~\text{TeV},\,\quad
M_1=500~\text{GeV},\quad M_2=1~\text{TeV},
\quad M_3=2.5~\text{TeV}, \nonumber\\[2mm]
&\quad A_t=A_b=A_\tau = 6.25~\text{TeV}\,.
\label{eq:alignmentscenario}
\end{eqnarray}

In order to obtain both alignment without decoupling and an acceptable
prediction for $\mh$ for {$\tan\beta\lesssim8$}, the parameters that
determine the stop masses take significantly larger values than in the
scenarios defined in sections \ref{sec:first scenario} and
\ref{sec:lightSUSY}. We also remark that the stop mixing parameter
$X_t = A_t -\mu\cot\beta$ is not fixed to a constant value over the
considered parameter space, and, due to the large value of $\mu$,
shows a significant dependence on $\tan\beta$. The trilinear coupling
$A_t$ is chosen in such a way that the prediction for $\mh$ is
maximized around the value of $\tan\beta$ for which the alignment
behavior occurs. In contrast, the specific choice of the gaugino mass
parameters $M_1$ and $M_2$ is rather irrelevant to the phenomenology
of this scenario, because the large value of $\mu$ implies a small
mixing between the lighter gaugino-like EW-inos and the heavier
higgsino-like EW-inos.  Hence, even if $M_1$ and $M_2$ were smaller
and Higgs-boson decays to EW-inos were kinematically allowed at lower
values of $\mA$, their rates would be strongly suppressed by small
couplings.

\begin{figure}[t]
\begin{center}
\includegraphics[width=0.75\textwidth]{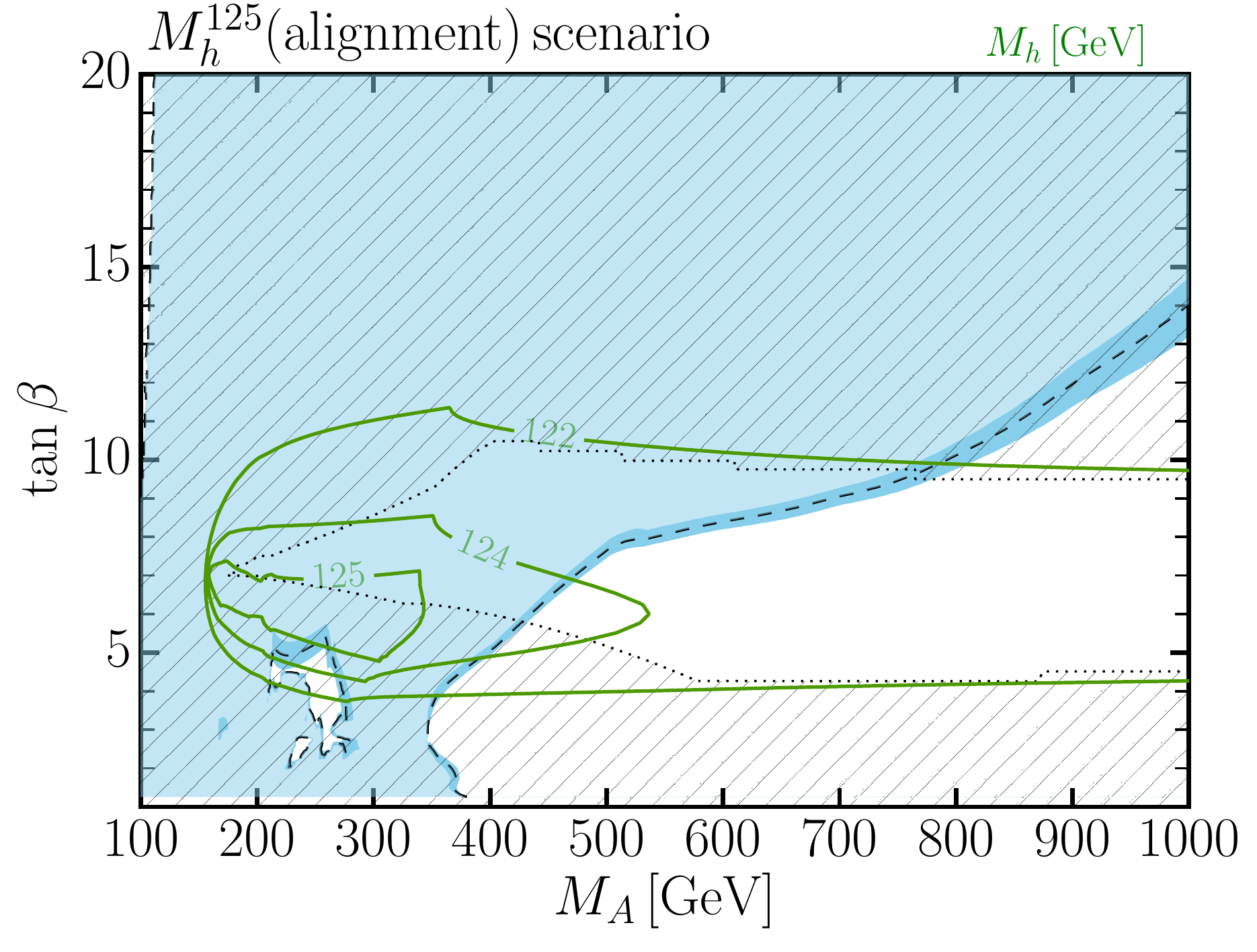}
\end{center}
\vspace*{-7mm}
\caption{Constraints on the \halignment\ scenario from Higgs searches at
  the LHC, in the $(\mA\,,\tb)$ plane. The green solid lines are
  predictions for the mass of the lighter \cp-even scalar $h$, the
  hatched area is excluded by a mismatch between the properties of $h$
  and those of the observed Higgs boson, and the blue area is excluded
  by the searches for additional Higgs bosons (the darker-blue band
  shows the theoretical uncertainty of the exclusion).}
\label{fig:halign_mhiggs}
\end{figure}

In \fig{fig:halign_mhiggs} we present the existing constraints on the
\halignment\ scenario from Higgs boson searches at the \lhc, in the
$(\mA\,,\tb)$ plane. To highlight the alignment effects, we
{focus} on the region with $100\,\text{GeV}\leq \mA \leq
1\,\text{TeV}$ and $1\leq\tan\beta\leq20$. The green solid lines show
that, as a result of the dependence of $X_t$ on $\tb$, the mass of the
lighter \cp-even scalar is maximized for $\tb\approx 6$. The decrease
of $\mh$ with increasing $\mA$, which contrasts with the usual
tree-level behavior, originates from two-loop diagrams involving
squarks and a heavy Higgs boson.\footnote{In view of the large
  hierarchy between the respective masses, a resummation of terms
  enhanced by powers of $\ln(M_S/\mA)$ may be necessary for
  $\mA\lesssim 500$~GeV, but that is not yet implemented in \FH.}

The shape of the hatched region ruled out by {\tt HiggsSignals} shows
that, in this scenario, the limit of alignment without decoupling is
realized {for $\tb \approx 7$}. Indeed, around this value of $\tb$
we see a wedge-shaped region in which the lighter \cp-even scalar can
be identified with the observed Higgs boson {for values of $\mA$
  as low as $170$~GeV}. When $\mA$ increases, the allowed region opens
up towards both smaller and larger values of $\tb$, as a consequence
of the usual decoupling behavior. However, even at large values of
$\mA$ the requirement that $\mh$ be compatible with the measured Higgs
mass (within the theoretical uncertainty of the MSSM prediction)
limits the allowed region {to the band in which
$4\lesssim\tb\lesssim10$}.

Finally, the blue region ruled out by the searches for heavy Higgs
bosons at the LHC, as determined by {\tt HiggsBounds}, is similar to
the corresponding regions in the \mhsc\ and \lstau\ scenarios, see
\figs{fig:mh125_mhiggs} and \ref{fig:mh125-ls_mhiggs}, respectively.
We see that a large portion of the wedge where alignment occurs
without decoupling is already excluded. However, we remark that the
lowest value of $\mA$ allowed in this scenario by both the heavy-Higgs
searches and the constraints on the properties of the lighter \cp-even
scalar, which is about $430$~GeV, is still significantly lower than
the bound of about $600$~GeV found in the first three scenarios.

\vspace*{-0.5mm}
\subsubsection{\boldmath{\mHsc} scenario}

MSSM scenarios in which the role of the SM-like Higgs boson is played
by the heavier \cp-even scalar $H$ are constrained by the fact that
the masses of all Higgs states are typically below $200$~GeV, thus the
searches for additional (neutral or charged) Higgs bosons at the LHC
already rule out most of the available parameter space. Earlier
proposals for such scenarios~\cite{Carena:2013ytb,Bechtle:2016kui}
focused on the region with $\mhp \approx m_t$ and low $\tan\beta$,
{triggered by} the fact that no search for charged Higgs bosons
had yet been performed in that region due to the lack of accurate
predictions for the signal, and that the region was not yet ruled out
by the searches for neutral Higgs bosons decaying to tau
pairs. Recently, the ATLAS collaboration published a search for
charged Higgs bosons~\cite{Aaboud:2018gjj} that covers the full range
of $90$~GeV~$\leq \mhp\leq 2$~TeV, employing the NLO predictions of
\citere{Degrande:2016hyf} for the total cross section for
charged-Higgs production with $\mhp \approx m_t$. However, in that
region the ATLAS search does not yet rule out low values of
$\tan\beta$. Moreover, the ATLAS search assumes that the charged Higgs
boson decays only through $H^\pm\rightarrow \tau^\pm\nu$, whereas in
scenarios where $H$ is SM-like the channel $H^\pm\rightarrow W^\pm h$
may become dominant when it is kinematically open.
It therefore seems worthwhile to devise a benchmark scenario in which
alignment without decoupling occurs for $\tb \approx 5$--$6$, and $H$
is approximately SM-like in the region that {has not been covered
  so far} by the charged-Higgs searches.

\bigskip

The \mHsc\ scenario is defined by the following input parameters:
\begin{eqnarray}
& M_{Q_3}=M_{U_3}=750\,\text{GeV} 
- 2\, ({M_{H^{\pm}}} -150\,\text{GeV})~,\nonumber\\[2mm]
& \mu = \left[5800\,\text{GeV} + 20 \, ({M_{H^{\pm}}} -150\,\text{GeV})\right]
\,M_{Q_3}/(750\,\text{GeV})~,\nonumber\\[2mm]
& A_t=A_b=A_\tau = 0.65 \,M_{Q_3},\quad M_{D_3}=M_{L_3}=M_{E_3}=2~\text{TeV}~,
\nonumber\\[2mm]
& M_1 = M_{Q_3} - 75\,\text{GeV}, 
\quad M_2=1\,\text{TeV},\quad M_3=2.5\,\text{TeV}~. 
\end{eqnarray}
To make the interplay with the charged-Higgs searches more
transparent, the free parameters in this scenario are chosen as $\tb$
and $\mhp$, with the latter varied between $150$ and $200$~GeV. In
contrast with the scenarios discussed earlier, the parameters that
determine the stop masses and couplings and the LSP mass are varied as
a function of $\mhp$, in order to maximize the experimentally viable
parameter space. In particular, the decrease of the stop masses with
increasing $\mhp$ has the purpose of delaying the crossing point at
which $h$ becomes SM-like and $\mH$ starts growing with $\mhp$. The
soft{-}SUSY-breaking bino mass $M_1$ is varied together with the
stop masses in order to ensure a compressed squark-neutralino
spectrum, still compatible with the existing monojet
searches~\cite{Aaboud:2017phn, Sirunyan:2017kiw} in the region of the
parameter space that is not excluded by other constraints. Finally,
the parameter $\mu$ takes particularly large values, about
{$8$--$9$} times the stop masses, to ensure that alignment occurs
at values of $\tb$ low enough to evade the bounds from
$H/A\rightarrow\tau^+\tau^-$ searches. Once again, we acknowledge the
possible tension with the theoretical bounds from the stability of the
scalar potential.

\begin{figure}[t]
\begin{center}
  \includegraphics[width=0.75\textwidth]{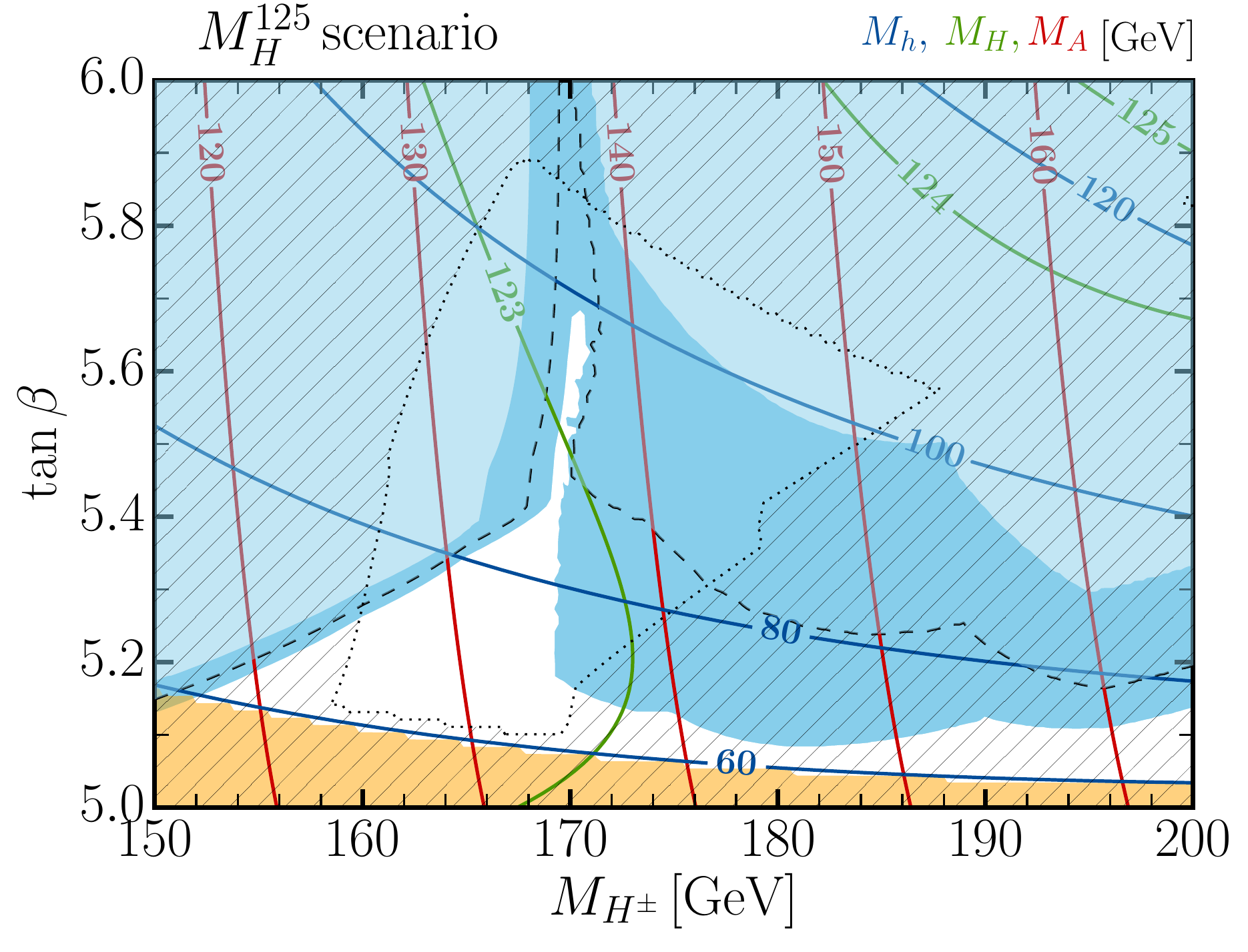}
\end{center}
\vspace*{-5mm}  
\caption{Constraints on the \mHsc\ scenario from Higgs searches at the
  LHC, in the $(\mhp,\tb)$ plane. The blue, green and red solid lines
  are predictions for the masses of $h$, $H$ and $A$,
  respectively. The hatched area is excluded by a mismatch between the
  properties of $H$ and those of the observed Higgs boson, and the
  areas bounded by dashed lines are excluded by the searches for
  additional Higgs bosons (the darker-blue {band shows} the
  theoretical uncertainty of the exclusion). At low $\tb$, the orange
  area is excluded by searches for $H\rightarrow hh$.}
\label{fig:mH125}
\end{figure}

In \fig{fig:mH125} we present the existing constraints on the
\mHsc\ scenario from Higgs searches at the LHC, in the relevant region
of the $(\mhp,\tb)$ plane. The blue, green and red solid lines
represent the masses of $h$, $H$ and $A$, respectively, and the
meaning of the hatched and blue regions is the same as in
\fig{fig:mh125_mhiggs}.  The figure shows that, although the
\mHsc\ scenario is already strongly constrained, a small region in
which $H$ can be identified with the observed Higgs boson is still
allowed. {In particular, the widest spread in $\tb$ is obtained
  for $\mhp \approx 170$~GeV, where $\tb$ ranges between $5.1$ and
  $5.8$, while the widest spread in $\mhp$ is obtained for $\tb
  \approx 5.3$, where $\mhp$ ranges between $160$~GeV and $175$~GeV}.

The non-hatched region allowed by {\tt HiggsSignals} is roughly shaped
as an irregular quadrilateral. We remark that this shape results from
a non-trivial interplay between the contributions of different Higgs
bosons to the signal strengths that {\tt HiggsSignals} compares with
the ATLAS and CMS measurements.\footnote{We stress again that these
  features depend to some extent on the way the ATLAS and CMS results
  are implemented in {\tt HiggsSignals}, and that dedicated analyses
  of the \mHsc\ scenario by the experimental collaborations would
  certainly yield more-accurate exclusion profiles.}
In parts of the allowed region, the decays $H\rightarrow b \bar b$ and
$H\rightarrow \tau^+\tau^-$ are somewhat suppressed with respect to
the SM prediction. However, since a mass resolution of $20$~GeV is
assumed for the corresponding measurements, the suppression can be
compensated for by the contribution of the \cp-odd scalar $A$ (this
explains the step-like feature in the border of the allowed region {at
$\mhp \approx 179$ GeV, where $\mA \approx 145$~GeV}). On the other
hand, when $\mh \gtrsim 105$~GeV the decays $h\rightarrow b \bar b$
and $h\rightarrow \tau^+\tau^-$ (as well as the associated production
of $h$ with top quarks) also start contributing to the signal
strengths, ruling out a region in which $H$ itself would be fairly
SM-like.
For decreasing $\tb$, radiative corrections to the \cp-even scalar
mass matrix that are enhanced by the large value of $\mu$ induce a
sharp decrease in the mass of the lighter eigenstate $h$. The lower
edge of the region allowed by {\tt HiggsSignals} lies where $\mh$
becomes lower than $\mH/2$ and the decay $H\rightarrow hh$ opens up,
suppressing the branching ratios of the SM-like decays of $H$. The
strip in which $\mh$ ranges between roughly $25$ and $60$ GeV is also
constrained by direct searches for the decays of the $125$-GeV Higgs
boson to pairs of light scalars. In particular, the region excluded by
the Run-1 CMS search for the decay channel $H\to hh\to b\bar{b}~
\mu^{\!+}\!\mu^-$~\cite{Khachatryan:2017mnf} is shown in orange in
\fig{fig:mH125}.

As shown by the areas bounded by dashed lines, the searches for
additional Higgs bosons implemented in {\tt HiggsBounds} cut parts of
the non-hatched region allowed by {\tt HiggsSignals}. In particular,
the region {with $\mhp \lesssim 170$~GeV} is excluded by the ATLAS
search for charged Higgs bosons decaying via $H^\pm\to
\tau^\pm\nu$~\cite{Aaboud:2018gjj}, down to values of $\tan\beta$ for
which the decay channel $H^\pm\to W^\pm h$ becomes dominant. The
region {with $\mhp \gtrsim 170$~GeV and $\tb \gtrsim 5.3\!-\!5.4$}
is instead excluded by the CMS searches for $A\rightarrow
\tau^+\tau^-$~\cite{CMS:2015mca,Sirunyan:2018zut}. However, the two
exclusion regions do not overlap, leaving a narrow unexcluded strip
{around $\mhp \approx 170$~GeV} which extends up to the edge of
the region allowed by {\tt HiggsSignals}.
We also note that the theoretical uncertainty of the exclusion region
at larger $\mhp$, shown as a darker-blue area in \fig{fig:mH125}, is
far from negligible{, because the rate for the process
  $pp\rightarrow A \rightarrow \tau^+\tau^-$ varies rather mildly over
  the considered range of $\tb$}. In particular, the use of the lowest
estimates for the production cross sections of the neutral Higgs
bosons would {significantly weaken} the bound from the
$A\rightarrow \tau^+\tau^-$ searches, extending the allowed region
{up to $\mhp \approx 185$~GeV}.

\begin{figure}[t]
\begin{center}
\includegraphics[width=0.49\textwidth]{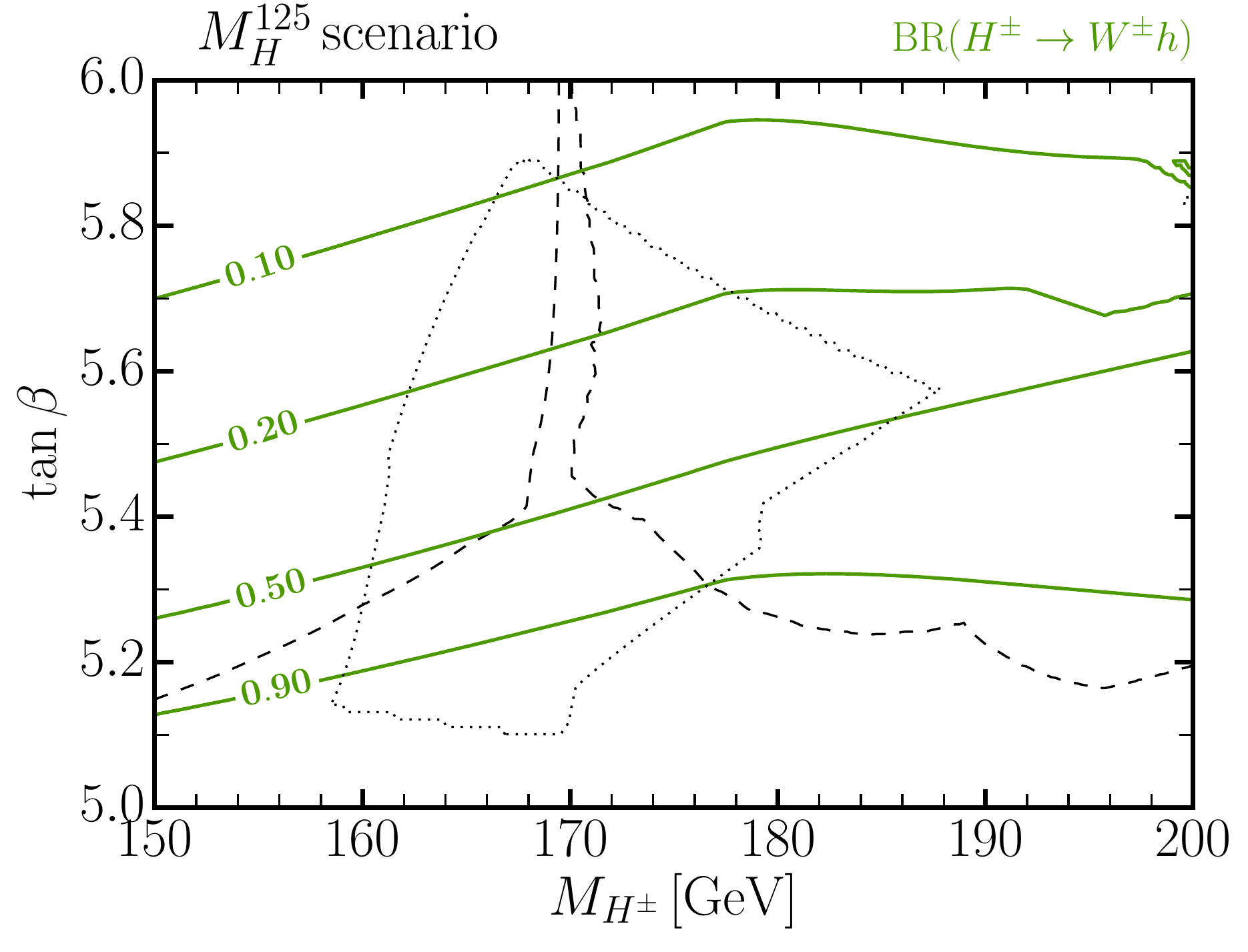}\hfill
\includegraphics[width=0.49\textwidth]{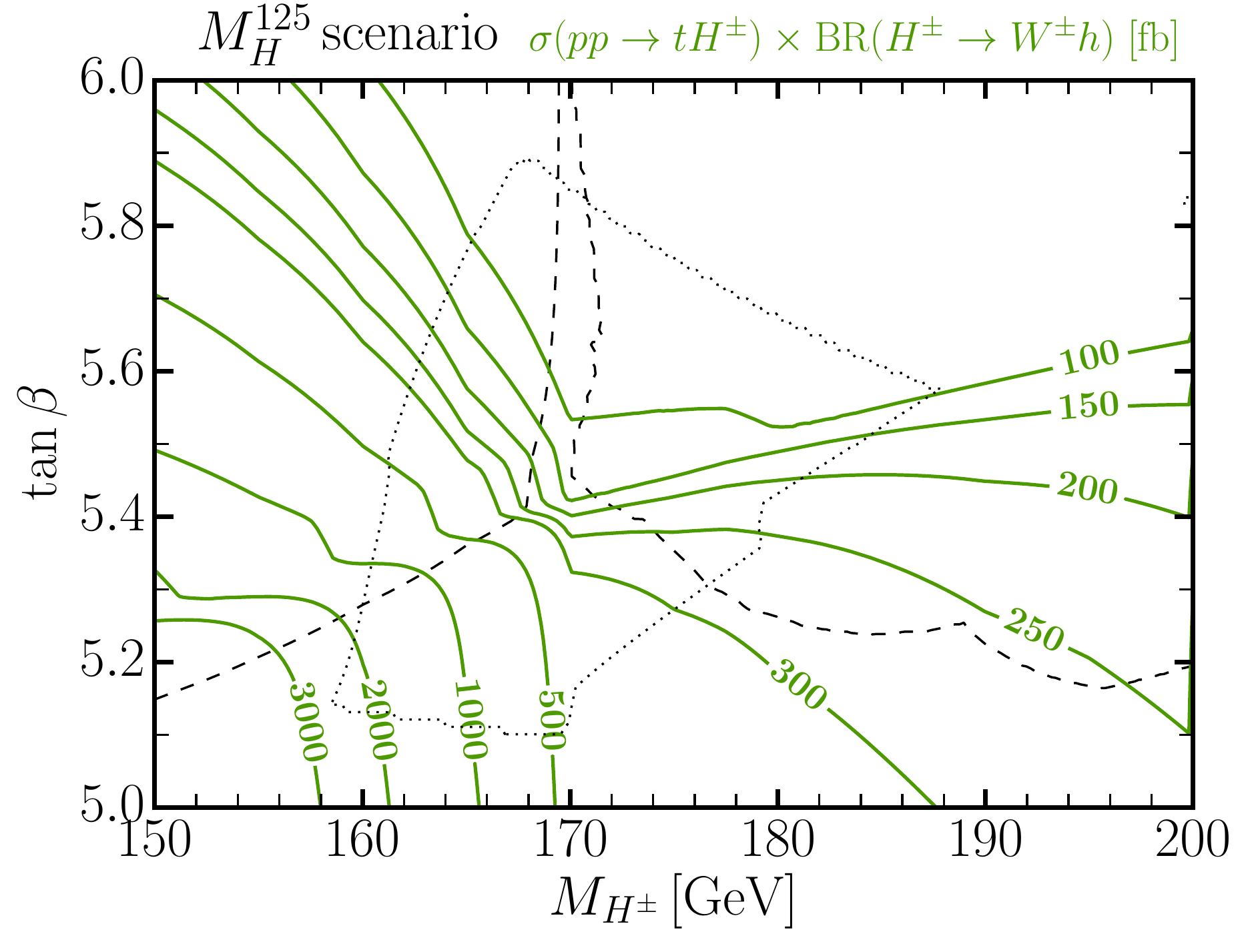}
\end{center}
\vspace*{-5mm}
\caption{Left: Branching ratio for the decay $H^\pm\rightarrow W^\pm
  h$ as a function of $\mhp$ and $\tb$ in the \mHsc\ scenario.  Right:
  Total rate (in fb) for the production of a charged Higgs boson in
  association with a top quark{, followed by the decay to a $W^\pm
    h$ pair,} at the LHC with $13$~TeV center-of-mass energy.  In each
  plot, the boundaries of the exclusion regions from \fig{fig:mH125}
  are shown as a dashed and a dotted black line, respectively.}
\label{fig:mH125-BRHp}
\end{figure}

Finally, we show in the left plot of \fig{fig:mH125-BRHp} the
branching ratio for the decay of a charged Higgs boson to a $W$ boson
and the {lighter} \cp-even scalar $h$ in the \mHsc\ scenario, and
in the right plot of \fig{fig:mH125-BRHp} the {total rate} (in fb)
for the production of a charged Higgs boson in association with a top
quark{, followed by the decay to a $W^\pm h$ pair,} at the LHC
with $13$ TeV center-of-mass energy {(a factor $2$ is included to
  account for both Higgs charges)}. The left plot shows that, in the
bulk of the allowed region {where $\tb \lesssim 5.4$}, the decay
channel $H^\pm \rightarrow W^\pm h$ is indeed the dominant one (with
the non-SM-like $h$ in turn decaying mostly to bottom quarks or
taus). The {right plot} suggests instead that, in this region, the
{total} rate for the process {$pp \to t H^\pm \to t (W^\pm
  h)$} {lies roughly between $200$~fb and $2$~pb}, making this
process an appealing candidate for a novel charged-Higgs search at the
LHC.



\subsection{\boldmath{\cpvint} scenario}
  
\label{sec:CPV}

Our sixth benchmark scenario, denoted as the ``\cpvint\ scenario'', is
characterized by \cp\ violation in the Higgs sector, and provides an
illustration of the interference effects in Higgs production and
decays discussed in section~\ref{sec:theory_interference}. The
relevant parameters are chosen such that the strongest interference
region is located near the exclusion contour of the corresponding
{scenario with real parameters}, the mass of the SM-like scalar is
near the observed value, and all EDMs are within the allowed ranges.
Our scenario can be considered a minimal \cp-violating setup, in that
the violation occurs as a result of a single non-zero {phase} for
the soft{-}SUSY-breaking Higgs-stop interaction term, defined as
$A_t \equiv |A_t|\,e^{i\phi_{A_t}}$. We choose a non-zero $\phi_{A_t}$
because this phase affects the Higgs sector already at one loop, via
the dominant corrections controlled by the large top Yukawa coupling,
while being the least constrained by EDMs. Indeed, the other relevant
phases\footnote{In fact, only the phases of various products of
  Lagrangian parameters are independent from each other. {With}
  appropriate field redefinitions, we can assume $M_2$ and the
  soft{-}SUSY-breaking Higgs mixing term $B_\mu$ to be real
  {without loss of generality}.}  $\phi_{A_f}$ (with $f\neq t$),
$\phi_{M_1}$, $\phi_{M_3}$ and $\phi_\mu$ induce smaller effects
and/or are more severely constrained (see, e.g.,
\citeres{Giudice:2005rz,Li:2010ax,Nakai:2016atk} for discussions of
EDM constraints in the \mssm).

In the \cpvint\ scenario the SUSY input parameters are fixed as
\begin{eqnarray}
& M_{Q_3}=M_{U_3}=M_{D_3}=M_{L_3}=M_{E_3}=2~\text{TeV}, \nonumber\\[2mm]
& \mu=1.65~\text{TeV},\,\quad M_1= M_2=1~\text{TeV},
\quad M_3=2.5~\text{TeV},\nonumber\\[2mm]
& |A_t|= \mu\cot\beta +2.8~\text{TeV} , \quad \phi_{A_t} = \frac{2\pi}{15},
\quad A_b= A_\tau=|A_t|~.
\label{par:cpv}
\end{eqnarray}

The non-zero phase $\phi_{A_t}$ leads to an admixture among the
neutral \cp-even scalars, $h$ and $H$, and the \cp-odd scalar, $A$,
into the loop-corrected mass eigenstates $h_1$, $h_2$ and $h_3$.
Since $A$ is not a mass eigenstate, the charged-Higgs mass \mhp\ is
used as a free input parameter together with $\tb$. In scans over the
$(\mhp,\tb)$ plane, $\tb$ can be varied between $1$ and $20$, and
$\mhp$ between $120$~GeV and $1$~TeV (to highlight the interference
effects, we focus on a smaller region of the parameter space than in
the scenarios of sections \ref{sec:first scenario} and
\ref{sec:lightSUSY}). As in the \mhsc\ scenario defined in section
\ref{sec:first scenario}, the mass parameters for all SUSY particles
are chosen to be {so large} that production and decays of the MSSM
Higgs bosons are only mildly affected by their presence. The largest
loop-induced SUSY effect arises through the $\Delta_b$ correction,
which takes on values similar to those in the \mhsc\ scenario.
We also remark that, in contrast to the scenarios defined in sections
\ref{sec:first scenario} and \ref{sec:lightSUSY}, the stop mixing
parameter $X_t = A_t -\mu\cot\beta$ is not fixed to a constant value
over the considered parameter space. However, for our choices of $A_t$
and $\mu$ the residual dependence of $X_t$ on $\tb$ has only a very
small effect on the stop masses when $\tan\beta\gtrsim 5$.

\begin{figure}[p]
\begin{center}
\includegraphics[width=0.66\textwidth]{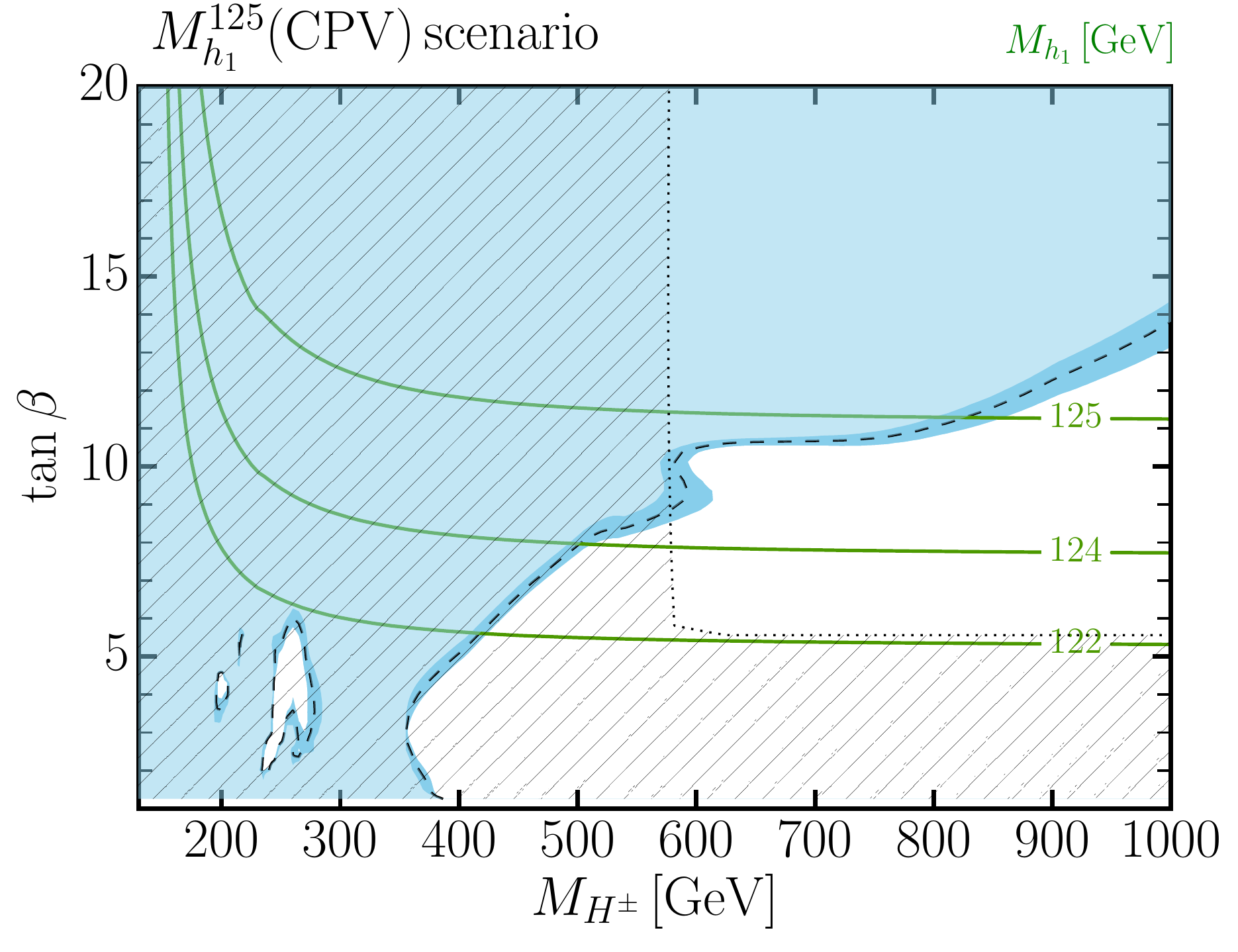} 
\end{center}
\vspace*{-8mm}
\caption{Constraints on the \cpvint ~scenario from Higgs searches at
  the LHC, in the $(\mhp\,,\tan\beta)$ plane.  The green solid lines
  are predictions for the mass of the lightest neutral scalar $h_1$,
  the hatched area is excluded by a mismatch between the properties of
  $h_1$ and those of the observed Higgs boson, and the blue area is
  excluded by the searches for additional Higgs bosons (the
  darker-blue band shows the theoretical uncertainty of the
  exclusion).}
  \label{fig:CPVint}
\end{figure}
\begin{figure}[p]
\begin{center}
\includegraphics[width=0.66\textwidth]{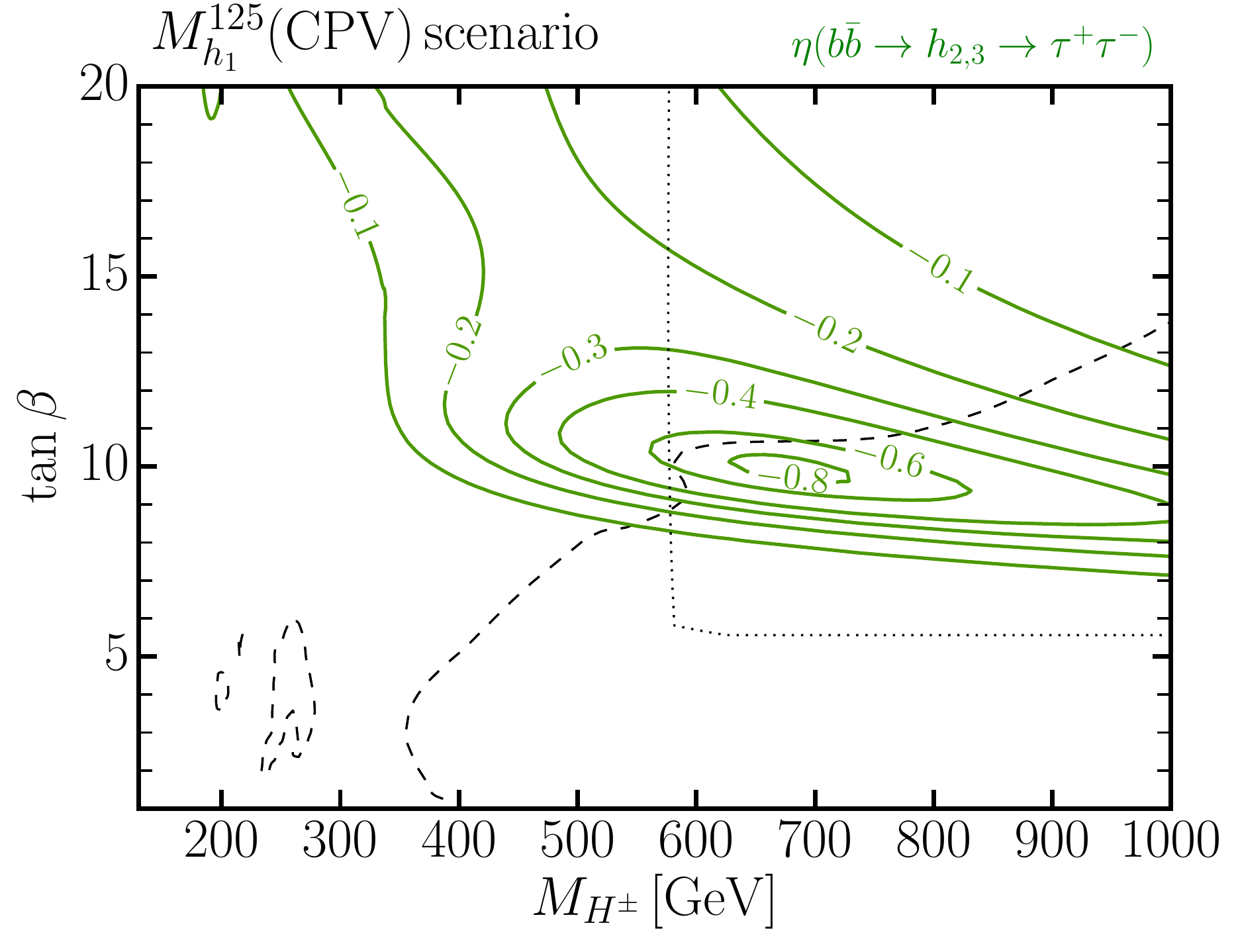}
\end{center}
\vspace*{-8mm}
\caption{The same exclusion boundaries as in \fig{fig:CPVint} above
  are overlaid with contours of the relative interference factor
  $\eta(b\bar b \to h_{2,3} \to \tau^+\tau^-)$ defined in
  \eqn{eq:eta}.}
  \label{fig:CPVint-eta}
\end{figure}

In \fig{fig:CPVint} we present, in the $(\mhp\,,\tb)$ plane, the
existing constraints on the \cpvint\ scenario from Higgs-boson
searches at the \lhc. The meaning of the blue and hatched exclusion
regions is the same as in \fig{fig:mh125_mhiggs}, but the green solid
lines refer here to the mass of the lightest mass eigenstate $h_1$.
The plot shows that $M_{h_1}$ takes on slightly larger values in the
\cpvint\ scenario than $\mh$ does in the \mhsc\ scenario, due to the
different choices for the soft{-}SUSY-breaking stop masses and for
$X_t$. The hatched region excluded by the requirement that the
properties of the lightest scalar match those of the observed Higgs
boson is qualitatively similar to the corresponding region in the
\mhsc\ scenario, see \fig{fig:mh125_mhiggs}. In contrast, the contour
of the blue region excluded by the LHC searches for the heavier
scalars $h_2$ and $h_3$ differs markedly from the corresponding
contour in \fig{fig:mh125_mhiggs}, with a ``bay''\!-shaped allowed
region opening up {around $\mhp\approx 650$~GeV and $\tb\approx
  10$}, where the most sensitive searches are the ones for $h_2/h_3
\rightarrow \tau^+\tau^-$.

To elucidate the origin of this feature, \fig{fig:CPVint-eta} shows
the relative interference factor for the production of the heavier
states $h_2$ and $h_3$ in bottom-quark annihilation followed by their
decay to a $\tau^+\tau^-$ pair, i.e.~the factor $\eta(b\bar b \to
h_{2,3} \to \tau^+\tau^-)$ defined in Eq.~(\ref{eq:eta}). The
interference factor for the gluon-fusion process -- which is very
similar to the one for bottom-quark annihilation -- has also been
taken into account in the evaluation of the exclusion region with {\tt
  HiggsBounds}.  However, the interference effect for bottom-quark
annihilation is the one with the strongest impact on the exclusion
contour, due to the larger cross section for intermediate to large
values of $\tan\beta$.  The plot clearly demonstrates that the
unexcluded ``bay'' corresponds to a region where a large destructive
interference reduces the prediction for {the} combined rate of
$h_2$ and $h_3$ to {less than about $30\%$} of the value that one
would obtain by neglecting the interference effects (in fact, the
minimal rate at the center of the interference region corresponds to
just $4\%$ of {the value without interference}). In this region
the mixing of the neutral \cp-even scalar $h$ with the heavier states
is negligible, while the mass eigenstates $h_2$ and $h_3$ are strong
admixtures of the \cp\ eigenstates $H$ and $A$, and they are almost
degenerate in mass.

The unconventional shape of the exclusion contour highlights the
importance of taking Higgs interference effects into account in
\cp-violating scenarios: even with the full luminosity of Run 2 of the
LHC, it is {possible} that ATLAS and CMS will {not} be able to
completely rule out {the region of the \cpvint\ scenario} in which
the heavy-Higgs rates {are} reduced by {up to} one order of
magnitude with respect to the corresponding rates in the
\cp-conserving case.
Moreover, we stress again that the occurrence of a sizable negative
interference is a {rather} universal feature of such \cp-violating
scenarios. {Even if the} region of the \cpvint\ scenario where the
strongest interference arises {should} eventually be ruled out by
tightening experimental constraints{, one should note that a
  slight change of parameters would} shift that region to larger
$H^\pm$ masses {and/or} lower values of $\tb$, while still
yielding $M_{h_1}$ near the observed value and fulfilling the EDM
constraints. At larger $\mhp$, viable interference regions require
larger $\phi_{A_t}$ in combination with {either} larger $\mu$
{or a larger constant term in the prescription for $|A_t|$, see
  \eqn{par:cpv}, and lower $\mu$.}  {At} lower $\tb${,} they
can {instead} be obtained by reducing $\phi_{A_t}$ and increasing
the constant term for $|A_t|$.

Finally, we checked with \FH\ and {\tt CPsuperH} that the relevant
interference region of the \cpvint\ scenario is compatible with the
bounds on the most sensitive EDMs, which in this case are the EDMs of
the electron and the neutron.  The strongest constraint arises from
the upper bound on the electron EDM~\cite{Baron:2013eja}, whereas in
case of a non-vanishing phase of $M_3$ the neutron
EDM~\cite{Baker:2006ts} would be very restrictive.  Due to the
enhancement of the electron EDM for larger $\tan\beta$ and smaller
Higgs masses, the upper{-}left corner of the $(\mhp,\,\tan\beta)$
plane in \fig{fig:CPVint} is in fact excluded by the electron EDM, but
only in a region that is also excluded by the direct searches for
heavy Higgs bosons.


\section{Conclusions}
\label{sec:conclusions}

In this paper we have proposed six new benchmark scenarios for MSSM
Higgs boson searches at the LHC. Our scenarios are expressed in terms
of TeV-scale parameters, chosen to illustrate different aspects of
Higgs phenomenology in the MSSM. They include one case with complex
parameters, but they all assume $R$-parity conservation and no flavor
mixing.  The scenarios are compatible -- at least over wide portions
of their parameter space -- with the most recent LHC results for the
Higgs-boson properties and the bounds on masses and couplings of new
particles. Each scenario contains one \cp-even scalar with mass around
$125$~GeV and SM-like couplings.  For each scenario we have
investigated the impact on the parameter space of the current
exclusion bounds from Higgs searches at LEP, the Tevatron and the LHC,
taking both experimental and theory uncertainties into account.

Our first scenario, ``\mhsc'', is characterized by relatively heavy
superparticles, so the Higgs phenomenology at the LHC resembles that
of a \thdm\ with MSSM-inspired Higgs couplings.  The second and third
scenario, ``\lstau'' and ``\lchi'', are characterized by some of the
superparticles -- staus and EW-inos, respectively -- being relatively
light. This affects the decays of the heavier Higgs bosons, weakening
the exclusion bounds from the $\tau^+\tau^-$ searches, as well as the
decay of the lighter \cp-even scalar to photons. On the other hand,
the possibility to look for additional Higgs bosons through their
decays to EW-inos opens up.  The fourth and fifth scenario are
characterized by the phenomenon of ``alignment without decoupling'',
in which one of the two neutral \cp-even scalars has SM-like couplings
independently of the mass spectrum of the remaining Higgs bosons.  In
particular, our fourth scenario, ``\halignment'', features a lighter
\cp-even scalar with SM-like couplings {for $\tb \approx 7$},
independently of the heavy Higgs-boson mass scale. It is thus
naturally in agreement with the measurements of the Higgs-boson
properties at the LHC also for relatively low values of $\mA$.  In the
fifth scenario, ``\mHsc'', the heavier \cp-even scalar is the one with
mass around $125$~GeV and SM-like couplings. In this scenario all
Higgs bosons are relatively light, posing very strong experimental
bounds on the parameter space. Nevertheless, the scenario appears to
be viable for $\tb \approx 5\!-\!6$ and very large values of~$\mu$,
and also features the novel signature of a relatively light charged
Higgs boson decaying to a $W$ boson and the lighter \cp-even
scalar. Finally, the sixth scenario, ``\cpvint'', incorporates
\cp\ violation in the Higgs sector and {gives rise to} a strong
admixture of the two heavier neutral states, leading to significant
interference effects in their production and decay which weaken the
exclusion bounds from $\tau^+\tau^-$ searches.

Spanning a wide variety of MSSM Higgs-boson phenomenology, we {hope}
that these new benchmark scenarios can serve as a guideline and
motivation for upcoming LHC searches for additional neutral and
charged Higgs bosons.


\section*{Note Added}

The recent improvement of the upper bound on the electron EDM of the
ACME collaboration~\cite{Andreev:2018ayy} excludes the phase
$\phi_{A_t}$ used in the \cpvint\ scenario, unless one invokes subtle
cancellations between different contributions to the EDM. However,
from our point of view such a scenario with two overlapping,
interfering Higgs-boson signals is still of interest. First, for
smaller phases compatible with newest EDM bounds the interference
effects are obviously smaller, but would still need to be taken into
account for a proper exclusion or discovery of heavier Higgs
bosons. It thus seems useful to study the relevance of such effects in
a scenario where they are clearly visible.  Second, such interferences
can be considered as a first step towards an implementation of
(experimentally non-resolvable) Higgs signal and background
interferences, which will be relevant in single-Higgs production as
well as in other channels (e.g.\ $t\bar t$, $Ah$ or $hh$) with
increasing integrated luminosity.

\section*{Acknowledgments}

This work was initiated in the context of the activities of the
\lhchxswg, and some of our findings were presented in a preliminary
form at recent meetings of the group. We thank F.~Staub for his help
concerning the usage of {\tt Vevacious}, {L.~Shang for
  communications about {\tt CheckMATE}}, and F.~Domingo,
{S.~Pa\ss{}ehr and J.~Wittbrodt} for {helpful} discussions.
E.~F.~is supported by the Minerva foundation.
The work of S.~H.~is supported in part by the MEINCOP Spain under
contract FPA2016-78022-P, in part by the Spanish ``Agencia Estatal de
Investigaci\'on'' (AEI) and the EU ``Fondo Europeo de Desarrollo
Regional'' (FEDER) through the project FPA2016-78022-P, and in part by
the AEI through the grant IFT Centro de Excelencia Severo Ochoa
SEV-2016-0597.
The work of P.~S.~is supported in part by the French ``Agence
Nationale de la Recherche'' (ANR), in the context of the LABEX ILP
(ANR-11-IDEX-0004-02, ANR-10-LABX-63) and of the grant
``HiggsAutomator'' (ANR-15-CE31-0002). P.~S.~also acknowledges support
by the Research Executive Agency (REA) of the European Commission
under the Initial Training Network ``HiggsTools''
(PITN-GA-2012-316704) and by the European Research Council (ERC) under
the Advanced Grant ``Higgs@LHC'' (ERC-2012-ADG\_20120216-321133).
T.~S.~acknowledges support from the DESY Fellowship programme.
{The work of C.~W.~at University of Chicago is supported in part by
U.S.~Department of Energy grant number DE-FG02-13ER41958, and his work
at ANL is supported in part by the U.S.~Department of Energy under
Contract No.~DE-AC02-06CH11357. C.~W.~would also like to thank the Aspen
Center for Physics, which is supported by National Science Foundation
grant PHY-1607611, for the kind hospitality during the completion of
this work.}
G.~W.~{acknowledges support} by Deutsche Forschungsgemeinschaft
through the SFB~676 ``Particles, Strings and the Early Universe''.


\vfill

{\footnotesize
\bibliographystyle{utphys}
\bibliography{MSSM_Run2_benchmarks}
}

\end{document}